\documentclass[final,1p,times]{elsarticle}

\usepackage{graphicx}

\usepackage{amssymb}

\usepackage{amsthm}

\usepackage[dvips]{epsfig,color}

\usepackage{lineno}

\newcommand{\PLB}[3]  {Phys.\ Lett.\ \textbf{B#1} (#2) #3}
\newcommand{\ZPC}[3]  {Z.\ Phys.\ \textbf{C#1} (#2) #3}

\newcommand{\EPA}[3]  {Eur.\ Phys.\ J.\ \textbf{A#1} (#2) #3}
\newcommand{\NIMA}[3] {Nucl.\ Instr.\ and Meth.\ \textbf{A#1} (#2) #3}

\newcommand{\PRL}[3]  {Phys.\ Rev.\ Lett.\ \textbf{#1} (#2) #3}

\newcommand{\CPC}[3]  {Comp.\ Phys.\ Comm.\ \textbf{#1} (#2) #3}
\newcommand{\EGeV}{\mbox{$\mathrm{E(GeV)}$}}
\newcommand{\epem}{\mbox{$\mathrm{e^+ e^-}$}}

\newcommand{\Zzero}{\mbox{${\mathrm{Z}}$}}
\newcommand{\Wboson}{\mbox{${\mathrm{W}}$}}

\newcommand{\WW} {\mbox{$\mathrm{W^+W^-}$}}

\newcommand{\qq}{\mbox{$\mathrm{q\overline{q}}$}}
\newcommand{\uu}{\mbox{$\mathrm{u\overline{u}}$}}

\newcommand{\uubar}{\mbox{$\mathrm{u\overline{u}}$}}
\newcommand{\ddbar}{\mbox{$\mathrm{d\overline{d}}$}}
\newcommand{\ssbar}{\mbox{$\mathrm{s\overline{s}}$}}
\newcommand{\qqp}{\mbox{$\mathrm{q'\overline{q}}$}}

\newcommand{\mZ}{\mbox{$M_{\mathrm{Z}}$}}

\newcommand{\EjjGeV}{\mbox{$E_{jj}(\mathrm{GeV})$}}
\newcommand{\Ejj}{\mbox{$E_{jj}$}}
\newcommand{\Ej}{\mbox{$E_{j}$}}

\newcommand{\OPAL}{\mbox{OPAL}}
\newcommand{\DELPHI}{\mbox{DELPHI}}
\newcommand{\ALEPH}{\mbox{ALEPH}}

\newcommand{\thetaqq}{\mbox{$\theta_{\mathrm{q\overline{q}}}$}}
\newcommand{\cosqq}{\mbox{$\cos\thetaqq$}}

\newcommand{\GeV}{\mbox{$\mathrm{GeV}$}}

\newcommand{\KL}{\mbox{$K_{\mathrm{L}}$}}
\newcommand{\KS}{\mbox{$K_{\mathrm{S}}$}}

\newcommand{\roots}{\mbox{$\sqrt{s}$}}

\newcommand{\rECAL}{R}
\newcommand{\rms}{\mbox{${\mathrm{rms}}_{90}$}}

\newcommand{\rmsn}{\mbox{${\mathrm{rms}}_{90}$}}

\newcommand{\PYTHIA}{\mbox{PYTHIA}}

\newcommand{\MARLIN}{\mbox{MARLIN}}
\newcommand{\MOKKA}{\mbox{MOKKA}}
\newcommand{\PANDORAPFA}{\mbox{PandoraPFA}}
\newcommand{\CALICE}{\mbox{CALICE}}
\newcommand{\GEANT}{\mbox{Geant4}}
\newcommand{\GEANTthree}{\mbox{Geant3}}

\newcommand{\LEPTRACKINGPROCESSOR}{\mbox{LEPTrackingProcessor}}
\newcommand{\SILICONTRACKING}{\mbox{SiliconTracking}}
\newcommand{\FULLLDCTRACKING}{\mbox{FullLDCTracking}}

\newcommand{\LCIO}{\mbox{LCIO}}

\def\etal{\mbox{{\it et al.}}}

\journal{Nuclear Instruments and Methods}

\begin{document}

\begin{flushright} 
CU-HEP-09/11 \\
July 2009
\end{flushright} 

\begin{frontmatter}


\title{Particle Flow Calorimetry and the \PANDORAPFA\ Algorithm}


\author{M. A. Thomson}
\ead{thomson@hep.phy.cam.ac.uk}

\address{Cavendish Laboratory, JJ Thomson Avenue, Cambridge CB3 0HE, United Kingdom.}

\begin{abstract}
The Particle Flow (PFlow) approach to calorimetry promises to deliver
unprecedented jet energy resolution for experiments at future high 
energy colliders such as the proposed International Linear Collider
(ILC). This paper describes the \PANDORAPFA\ particle flow algorithm
which is then used to perform the first systematic study of
the potential of high granularity PFlow calorimetry. For simulated
events in the ILD detector concept, a jet energy resolution of 
$\sigma_E/E \lesssim 3.8\,\%$ is achieved for $40-400$\,GeV jets.
This result, which demonstrates that high granularity PFlow calorimetry 
can meet the challenging ILC jet energy resolution goals, 
does not depend strongly on the details of the Monte Carlo modelling of 
hadronic showers. The \PANDORAPFA\ algorithm is also used to investigate 
the general features of a collider detector optimised
for high granularity PFlow calorimetry. Finally, a first study of 
the potential of high granularity PFlow calorimetry at a multi-TeV lepton collider, 
such as CLIC, is presented.  
\end{abstract}


\end{frontmatter}

\section{Introduction}
\label{Introduction}

In recent years the concept of high granularity 
Particle Flow calorimetry~\citep{bib:pfa} 
has been developed in the context of the proposed International Linear 
Collider (ILC).
Many of the interesting physics processes at the ILC~\citep{bib:ILC} will be characterised by
multi-jet final states, often accompanied by charged leptons 
and/or missing transverse energy associated with neutrinos or the 
lightest super-symmetric particles. The reconstruction of the 
invariant masses of two or more jets will provide a powerful tool 
for event reconstruction and event identification. Unlike at LEP, where 
kinematic fitting~\citep{bib:mwfit} enabled precise invariant mass 
reconstruction, at the ILC di-jet mass reconstruction will rely on
the jet energy resolution of the detector. 
The goal for jet energy resolution at the ILC is that it 
is sufficient to cleanly separate $\Wboson$ and $\Zzero$ 
hadronic decays. An invariant
mass resolution comparable 
to the gauge boson widths, {\it i.e.} $\sigma_m/m = 2.7\,\% \approx \Gamma_W/m_W 
\approx \Gamma_Z/m_Z$, leads to an effective $3.6\,\sigma$ separation 
of the $\mathrm{W}\rightarrow\qqp$ and $\mathrm{Z}\rightarrow\qq$ 
mass peaks, {\em i.e.} the optimal invariant mass cut corresponds to
$+1.8\,\sigma$ ($-1.8\,\sigma$) in the reconstructed
$\Wboson$ $(\Zzero)$ mass distributions. 

In the traditional calorimetric approach, the jet energy is obtained 
from the sum of the energies deposited in the electromagnetic and 
hadronic calorimeters (ECAL and HCAL). 
This typically results in a jet energy resolution of the form
\begin{eqnarray}
    \frac{\sigma_E}{E} &=&  \frac{\alpha}{\sqrt{\EGeV}} \oplus \beta.  \label{eqn:cal} 
\end{eqnarray}
The stochastic term, $\alpha$, is usually greater than $\sim$60\,\%. The
constant term, $\beta$, which encompasses a number of effects, is typically
a few per cent. For high energy jets there also will be a contribution 
from the non-containment of the hadronic showers. 
The stochastic term in the jet energy resolution results in a 
contribution to the di-jet mass resolution of $\sigma_m/m \approx \alpha/\sqrt{E_{jj}}$, where $E_{jj}$ 
is the energy of the di-jet system in GeV. At the ILC, operating at 
centre-of-mass energies $\roots = 0.5-1.0$\,TeV, the 
typical di-jet energies for interesting physics processes 
will be in the range $150-350$\,GeV. 
Hence to achieve the ILC goal of
$\sigma_m/m = 2.7$\,\%, the
stochastic term must be $\lesssim 30\,\%/\sqrt{\EGeV}$. This is 
unlikely to be achievable with a traditional approach to calorimetry. 

\subsection{The Particle Flow Approach to Calorimetry}

Measurements of jet fragmentation at LEP have provided detailed 
information on the particle composition of jets 
({\it e.g.}~\citep{bib:Knowles,bib:Green}).  On average, after the 
decay of short-lived particles, roughly 62\% of the jet energy  
is carried by charged particles (mainly hadrons), around 27\% 
by photons, about 10\% by long-lived neutral 
hadrons ({\em e.g.} $n$, $\overline{n}$ and $\KL$), and around 1.5\,\% by neutrinos.
Hence, approximately 72\,\% of the jet energy is measured in the HCAL and
the jet energy resolution is limited by
the relatively poor HCAL energy resolution, typically $\gtrsim 55\,\%/\sqrt{\EGeV}$. 
The LEP collaborations, most notably ALEPH, and other collider experiments ({\em e.g.}
H1, D0 and CMS) have obtained improved jet energy resolution using the 
Energy Flow~\citep{bib:Aleph-jet} approach, whereby energy deposits in the 
calorimeters are removed according to the momentum of the charged particle tracks.
Using this method, ALEPH achieved a jet energy resolution (for $\roots = \mZ$) equivalent to
$\sigma_E/E \approx 65\,\%/\sqrt{E(\GeV)}$~\citep{bib:Aleph-jet}. This is 
the best jet energy resolution of the four LEP experiments, but is
roughly a factor two worse than required for the ILC. 

It is widely believed that the most promising strategy\footnote{The only alternative proposed to date
is that of Dual Readout calorimetry as 
studied by the DREAM collaboration~\cite{bib:dual}.} for 
achieving the ILC
jet energy goal is the Particle Flow (PFlow) approach to calorimetry. This 
extends the concept of Energy Flow to a highly granular detector. 
In contrast to a purely calorimetric measurement, PFlow calorimetry requires the 
reconstruction of the four-vectors of all visible particles in an event. The 
reconstructed jet energy is the sum of the energies of the individual 
particles. The momenta of charged particles are measured in the tracking 
detectors, while the energy measurements for photons and neutral hadrons 
are obtained from the calorimeters. In this manner, the HCAL is used to
measure only $\sim10\,\%$ of the energy in the jet. 
If one were to assume calorimeter resolutions of $\sigma_E/E = 0.15/\sqrt{\EGeV}$
for photons and $\sigma_E/E = 0.55\sqrt{\EGeV}$ for hadrons, 
a jet energy resolution of $0.19/\sqrt{\EGeV}$ would be obtained with the 
contributions from tracks, photons and neutral hadrons as given in 
Table~\ref{tab:res}. In practice, this level 
of performance can not be achieved as it is not possible to perfectly 
associate all energy deposits with the correct particles. For example, 
if the calorimeter hits from a photon are not resolved from a charged 
hadron shower, the photon energy is not accounted for. Similarly, if part 
of charged hadron shower is identified as a separate cluster, the energy 
is effectively double-counted as it is already accounted for by the
track momentum. This {\em confusion} rather than calorimetric
performance is the limiting factor in PFlow calorimetry. 
Thus, the crucial aspect of PFlow calorimetry is the ability to correctly assign 
calorimeter energy deposits to the correct reconstructed 
particles. This places stringent requirements on the granularity of 
the ECAL and HCAL. From the point of view of 
event reconstruction, the sum of calorimeter energies is replaced
by a complex pattern recognition problem, namely the Particle Flow reconstruction
Algorithm (PFA). The jet energy resolution obtained is a combination 
of the intrinsic detector performance and the performance of the PFA software.

\begin{table*}[th]
\renewcommand{\arraystretch}{1.2}
\begin{center}
\begin{tabular}{|l|cccc|}
\hline
{\bf Component}          & {\bf Detector} & {\bf Energy Fract.} & {\bf Energy Res.} & {\bf Jet Energy Res.} \\ \hline
Charged Particles ($X^\pm$) & Tracker           & $\sim0.6\,\Ej$ & $10^{-4}\,E^2_{X^\pm}$  & 
   $<3.6\times10^{-5}\,E^2_{j}$ \\
Photons $(\gamma)$          & ECAL              & $\sim0.3\,\Ej$ & $0.15\,\sqrt{E_\gamma}$ & 
   $0.08\,\sqrt{\Ej}$ \\
Neutral Hadrons $(h^0)$     & HCAL              & $\sim0.1\,\Ej$ & $0.55\,\sqrt{E_{h^0}}$  &
$0.17\,\sqrt{\Ej}$ \\ \hline 
\end{tabular}
\caption{Contributions from the different particle components to the jet-energy resolution
 (all energies in GeV). The table lists the approximate fractions of charged particles, 
 photons and neutral hadrons in a jet of energy, $\Ej$, and the assumed single particle 
 energy resolution. \label{tab:res}}
\end{center}
\renewcommand{\arraystretch}{1.0}
\end{table*}

The \PANDORAPFA\ algorithm was developed to study PFlow calorimetry
at the ILC. \PANDORAPFA\ is a {C++} implementation of a PFA 
running in the \MARLIN~\citep{bib:marlin} reconstruction framework. 
It was developed and optimised 
using simulated physics events generated with the \MOKKA~\citep{bib:mokka} 
program, which provides a detailed \GEANT~\citep{bib:geant4} simulation of potential 
detector concepts for the ILC. In particular, \PANDORAPFA\ was developed using
the \MOKKA\ simulation of the LDC~\citep{bib:ldc} detector concept and, more recently,
the ILD~\citep{bib:ildloi} detector concept. The algorithm is designed to be 
sufficiently flexible to allow studies of PFlow for different detector
designs. Whilst a number of PFAs~\cite{bib:pfaMorg,bib:pfaWolf,bib:pfaSid} 
have been developed for the ILC, \PANDORAPFA\ is the most sophisticated and
best performing algorithm. 
In this paper \PANDORAPFA\ is described in detail. It is then
used to study the potential at a future high energy lepton collider 
of PFlow calorimetry with a highly granular detector, in this case the ILD 
detector concept. 

\section{Overview of the ILD Detector Model}

\label{sec:ild}

\begin{figure}[hbt]
\epsfxsize=1.0\columnwidth
\centerline{\includegraphics[width=1.0\columnwidth]{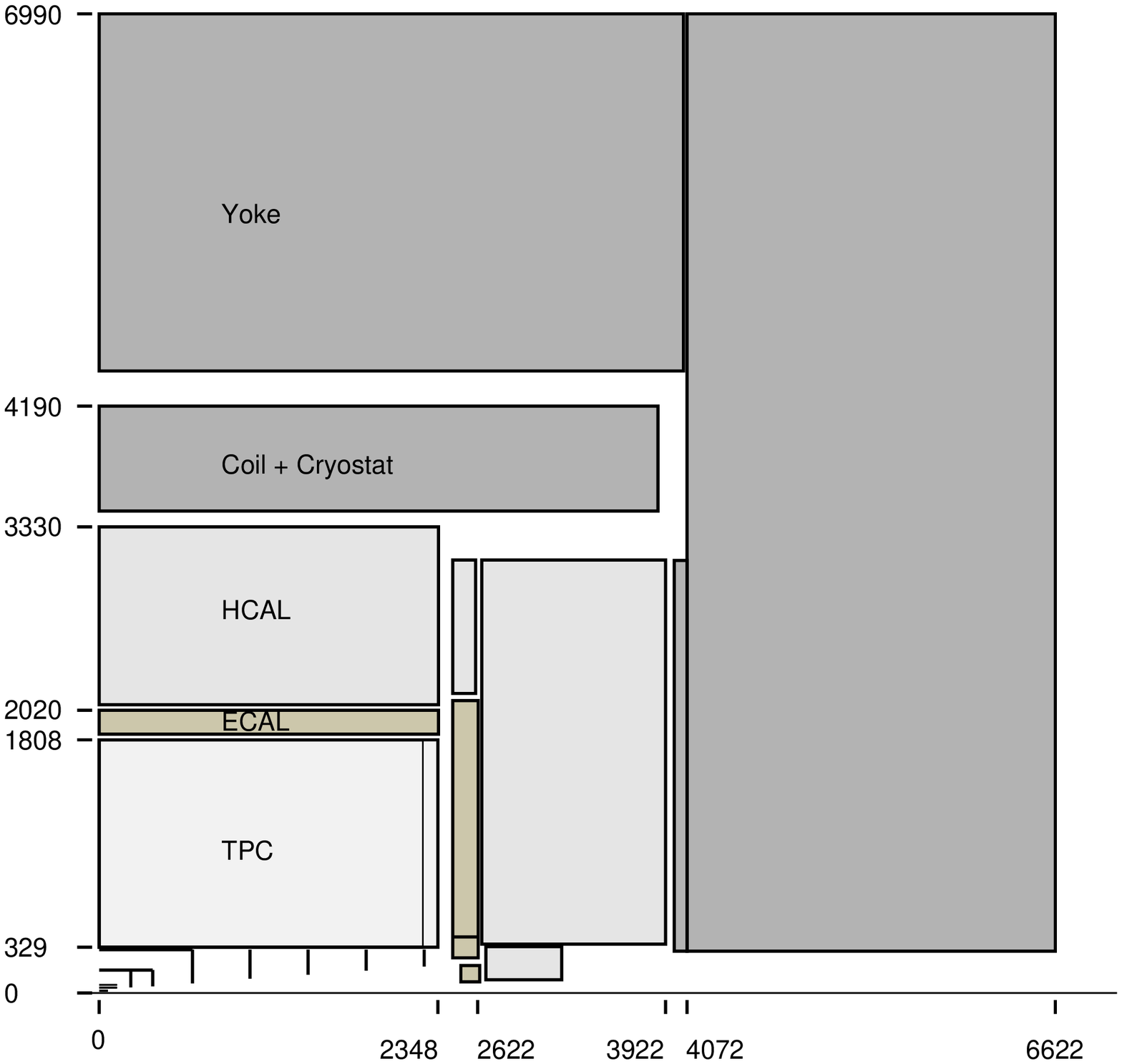}}
\caption{A quadrant of the ILD detector concept showing the main dimensions and layout of the
          sub-detector components.
\label{fig:ildquad}}
\end{figure}

The ILD detector concept~\cite{bib:ildloi}, 
shown in Figure~\ref{fig:ildquad}, consists of a vertex detector, tracking detectors, 
ECAL, HCAL and muon chambers. It represents a possible configuration of a
detector suitable for PFlow calorimetry. Specifically, for 
the ECAL and HCAL the emphasis is on granularity, 
both longitudinal and transverse, rather than solely energy resolution. Suitable
candidate technologies are being studied by the \CALICE\ 
(calorimetry for the ILC) collaboration~\cite{bib:calice}. Amongst these
are the Silicon-Tungsten ECAL and Steel-Scintillator HCAL designs assumed for
the baseline ILD detector simulation.

Both the ECAL and HCAL are located inside a solenoid which is taken 
to produce the 3.5\,T magnetic field. 
The main tracking detector is simulated
as a time projection chamber (TPC) with an active gas 
volume of half-length 2.25\,m and inner and outer radii of 0.39\,m 
and 1.74\,m respectively. The vertex detector consists of 6 layers of Silicon with an inner radius
of 15\,mm from the interaction point (IP). 
The tracking is complemented by two barrel Silicon strip detectors
between the vertex detector and the TPC and seven Silicon forward tracking disks. 
The ECAL is simulated as a Silicon-Tungsten sampling calorimeter 
consisting of 29 layers. The first 20 layers 
have 2.1\,mm thick Tungsten and the last 9 layers have 4.2\,mm thick Tungsten. 
The high resistivity Silicon is segmented into $5\times\,5\,\mathrm{mm}^2$ pixels.
At normal incidence, the ECAL corresponds to 23 radiation lengths  ($X_0$)  and 0.8 nuclear
interaction lengths  ($\lambda_I$). The HCAL is simulated as a Steel-Scintillator 
sampling calorimeter comprising 48 layers of 20\,mm thick Steel and 5\,mm thick
$3\times\,3\,\mathrm{cm}^2$ plastic scintillator tiles. At normal incidence the 
HCAL is 6\,$\lambda_I$ thick. 

The ECAL and HCAL in the ILD concept are well matched to the
requirements of PFlow calorimetry. 
Tungsten is the ideal absorber material for the
ECAL; it has a short radiation length and small Moli\`{e}re radius (see Table~\ref{tab:properties}) 
which leads
to compact electromagnetic (EM) showers. It also has a large ratio of interaction
length to radiation length which means that hadronic showers will tend to be longitudinally
well separated from EM showers. The   
$5\times\,5\,\mathrm{mm}^2$ transverse segmentation takes full advantage
of the small Moli\`{e}re radius. Steel is chosen as the HCAL absorber, primarily
for its structural properties. The $3\times\,3\,\mathrm{cm}^2$ HCAL transverse segmentation
is believed to be well matched to the requirements of PFlow calorimetry (see Section~\ref{sec:calseg}).

\begin{table}[h]
\begin{center}
\begin{tabular}{|c|cccc|}
\hline
\textbf{Material} & \textbf{\boldmath$\lambda_I$/cm} & \textbf{\boldmath$X_0$/cm}
 & \textbf{\boldmath$\rho_\mathrm{M}$/cm} & \textbf{\boldmath$\lambda_I/X_0$}  \\ \hline
Fe      & 16.8 & 1.76 & 1.69 &  9.5 \\
Cu      & 15.1 & 1.43 & 1.52 & 10.6 \\
W       &  9.6 & 0.35 & 0.93 & 27.4 \\
Pb      & 17.1 & 0.56 & 1.00 & 30.5 \\ \hline
\end{tabular}
\caption{Comparison of interaction length, $\lambda_I$, radiation length, 
$X_0$,
and Moli\`{e}re radius, $\rho_\mathrm{M}$, for Iron, Copper, Tungsten and Lead. Also given is the
ratio of $\lambda_I/X_0$.
\label{tab:properties}}
\end{center}
\end{table}

\section{Reconstruction Framework}

The performance of PFlow calorimetry depends strongly on the reconstruction 
software. For the results obtained to be meaningful, it is essential that both the
detector simulation and the reconstruction chain are as realistic as possible.  
For this reason no Monte Carlo (MC) information is used at
any stage in the reconstruction as this is likely to lead to an overly-optimistic 
evaluation of the potential performance of PFlow calorimetry.

\PANDORAPFA\ runs in the \MARLIN~\citep{bib:marlin} {C++}
framework developed for the LDC and ILD detector 
concepts. The input to \PANDORAPFA\ (in \LCIO~\cite{bib:lcio} format) is a list  
of digitised hits in the calorimeters and a list of reconstructed tracks.
Tracks in the TPC are reconstructed using
a \MARLIN\ processor, \LEPTRACKINGPROCESSOR, 
adapted from the TPC pattern recognition software~\citep{bib:tracking} based on
that used by \ALEPH\
and track fitting software used by \DELPHI~\citep{bib:delphitracking}. 
Reconstruction of tracks in the inner Silicon detectors is performed by a custom 
processor, \SILICONTRACKING~\citep{bib:ldctrackingprocessor}. TPC and Silicon
track segments are combined in a final tracking processor, 
\FULLLDCTRACKING~\citep{bib:ldctrackingprocessor}. 

\begin{figure}[hbt]
\epsfxsize=1.0\columnwidth
\centerline{\includegraphics[width=1.0\columnwidth]{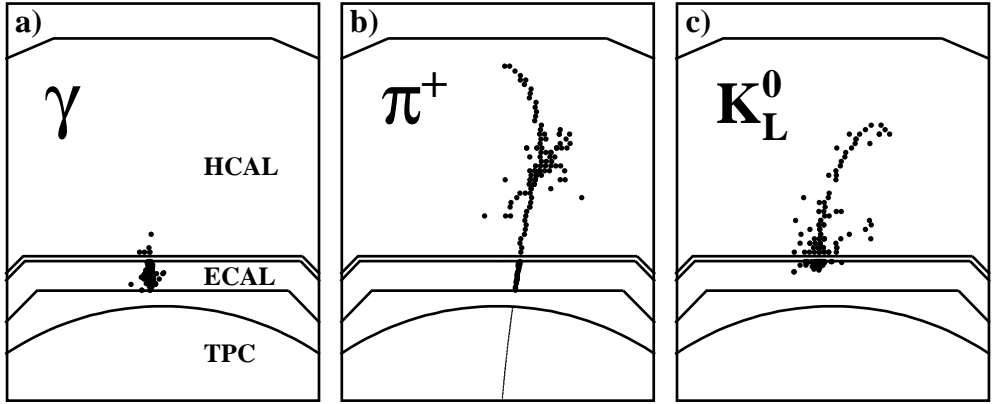}}
\caption{ Example simulated single particle interactions in the ILD
          detector concept: a) a 10 GeV photon; b) a 10 GeV $\pi^+$ and
          c) a 10 GeV \KL. Hits in the TPC, ECAL and HCAL are shown. For the
          ECAL (HCAL) all hits with energy depositions $>0.5\, (0.3)$ minimum 
          ionising particle equivalent are displayed. Simulated TPC hits are 
          digitised assuming 227 radial rows of readout pads.}
\label{fig:exampleParticles}
\end{figure}

\PANDORAPFA\
combines the tracking information with hits in the high granularity
calorimeters to reconstruct the individual particles in the event. As an
example of the information used in the reconstruction, Figure~\ref{fig:exampleParticles}
shows a photon, a charged hadron ($\pi^+$) and a neutral hadron (\KL) as simulated
in the ILD detector concept.

\section{The \PANDORAPFA\ Particle Flow Algorithm}

The \PANDORAPFA\ algorithm performs calorimeter clustering and 
PFlow reconstruction in eight main stages: 
 {\it {1) Track Selection/Topology:}} track topologies such as kinks and decays of neutral
          particles in the detector volume ({\em e.g} $\KS\rightarrow\pi^+\pi^-$) are identified.
 {\it {2) Calorimeter Hit Selection and Ordering:}} isolated hits, defined on the basis of proximity 
          to other hits, are removed from the initial clustering stage. 
          The remaining hits are ordered into {\it pseudo-layers} and information related
          to the geometry and the surrounding hits are stored for use in the reconstruction. 
{\it {3) Clustering:}} the main clustering algorithm is a cone-based forward projective 
          method~\cite{bib:thomsonlcws07} working 
          from innermost to outermost pseudo-layer. The algorithm starts by {seeding} clusters
          using the projections of reconstructed tracks onto the front face of the
          ECAL.  
{\it {3A) Photon Clustering:}}  
      \PANDORAPFA\ can be run in a mode where the above clustering algorithm 
       is performed in two stages. In the first stage, only ECAL hits are considered with the aim of 
       identifying energy deposits from photons. 
       In the second stage the clustering algorithm
       is applied to the remaining hits.
{\it {4) Topological Cluster Merging:}} by design the initial clustering stage errs on the side of 
      splitting up true clusters rather than merging energy deposits from more than one particle into a
      single cluster. Clusters are then combined on the basis of clear topological signatures in the high
      granularity calorimeters. The topological cluster merging algorithms are only applied 
      to clusters which have not been identified as photons. 
{\it {5) Statistical Re-clustering:}} The previous four stages of the algorithm are found to 
      perform well for jets with energies of less than $50$\,GeV. 
      For higher energy jets the performance degrades due to
      the increasing overlap between hadronic showers from different particles. Clusters which
      are likely to have been created from the merging of hits in showers from more than one particle 
      are identified on the basis of the compatibility of the cluster energy, $E_C$, and the
      associated track momentum, $p$. In the case of an inconsistent energy-momentum match, attempts
      are made to re-cluster the hits by re-applying the clustering algorithm 
      with different parameters, until the cluster splits to give a 
      cluster energy consistent with the momentum of the associated track.
{\it {6) Photon Recovery and Identification:}} 
      A more sophisticated, shower-profile based, photon-identification algorithm is then 
      applied to the clusters, improving the tagging of photons. It is also used to  recover
      cases where a primary photon is merged with a hadronic shower from a charged particle.
{\it {7) Fragment Removal:}} ``neutral clusters'' which are {\em fragments} 
       of charged particle hadronic showers are identified.
{\it {8) Formation of Particle Flow Objects:}} 
     The final stage of the algorithm is to create the list of reconstructed particles,
      Particle Flow Objects (PFOs), and associated four-momenta. 

The essential features of each of the above stages are described in more detail
below. The description includes the main configuration parameters 
which determine the behaviour of the algorithms. These can be defined at runtime. The 
default values, which are optimised for the ILD concept, are given.

\subsection{Track selection/topology}

Tracks are projected onto the front face of the 
ECAL using a helical fit to the last 50 hits on the 
reconstructed track (no account is taken for energy loss along the 
trajectory in the TPC gas). Tracks are then
classified according to their likely origin. For example, neutral particle 
decays resulting in two charged particle tracks ($V^0$s) are
identified by searching for pairs of tracks which 
are consistent with coming from a single point displaced from the 
IP. Charged particle decays to a single charged particle and any 
number of neutral particles (kinks) are identified on the basis of the distance of closest
approach of the parent and daughter tracks. Similarly, interactions in the tracking volume
(prongs) are identified. This information, along with the original track parameters 
and the projection of the track onto the front face of the ECAL, is stored in 
{\tt ExtendedTrack} objects for use in the subsequent event reconstruction.
 
\subsection{Calorimeter Hit Selection and Ordering} 

In addition to the reconstructed tracks, the input to \PANDORAPFA\ is a list 
of digitised calorimeter hits. For each hit, the position $(x,y,z)$, the energy deposition, 
and the physical layer in the ECAL/HCAL are specified. Based on this information, 
{\tt ExtendedCaloHit} objects are formed. These hits are self-describing and 
incorporate information relating to both the geometry of the detector (accessed from the
GEAR\citep{bib:gear} geometry description) 
and information related to the density of calorimeter hits in the neighbouring region.
The five main steps in the calorimeter hit processing  
(calibration, geometry, isolation, MIP identification, ordering) 
are described below.

\subsubsection{Calibration}
The energy of each calorimeter hit is converted to a minimum ionising 
particle (MIP) equivalent (at normal incidence) using a calibration 
factor {\tt CALMIPcalibration}. 
Different calibration factors are used for ECAL and HCAL hits. 
Hits are only retained if they are above a MIP-equivalent threshold
of {\tt CalMIPThreshold}  (with default values of
[0.5] and [0.3] for ECAL and HCAL respectively). 
The MIP equivalent energy deposit is then converted into calorimetric measurement
using  MIP to GeV calibration factors, {\tt CalMIPToGeV}, for the 
ECAL and HCAL.
In general, the calorimeters will not be compensating, and 
separate energy measurements are calculated for the hypotheses that the hit is either
part of an EM or hadronic shower. The final choice of which 
energy to use depends on the whether the shower to which a hit is associated is
ultimately identified as being EM in nature. To allow for 
calorimeters with different absorber thicknesses as a function of depth, 
the calibration factor applied is proportional to the absorber thickness of the layer 
in front of the hit. Initial values for the calibration factors are determined from 
MC samples of single muons, photons and $\KL$s. The muon sample is used to determine
the MIP calibration, the photon sample is used to determine the ECAL calibrations and
the $\KL$s are used to determine the initial HCAL calibration. Since the neutral
hadrons in jets are a mixture of $\KL$s, neutrons and anti-neutrons, the initial
HCAL calibration is modified (typically by $\sim$5\,\%) on the basis of 
minimising the jet energy resolutions for MC samples of jets. 
A single set of calibration factors is used for the subsequent studies. 

\subsubsection{Geometry information}
\begin{figure}[hbtp]
\epsfxsize=1.0\columnwidth
\centerline{\includegraphics[width=1.00\columnwidth]{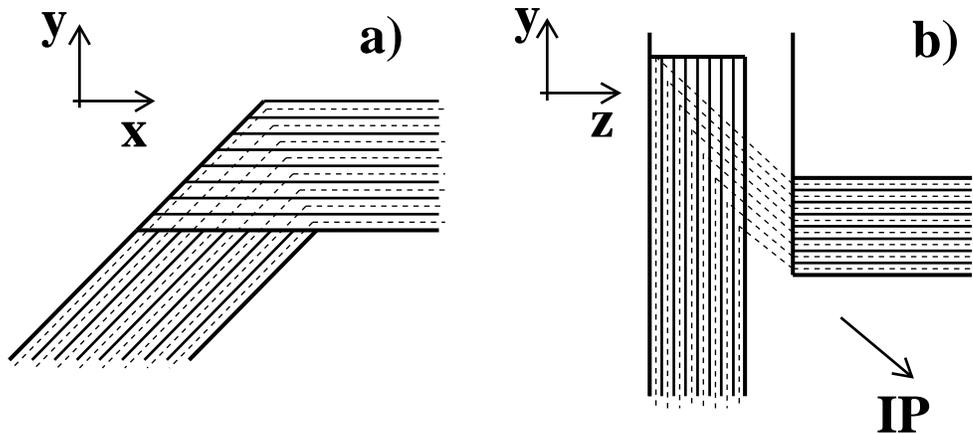}}
\caption{Schematic showing the definition of the pseudo-layer assignment for
         calorimeter hits. The solid lines indicate the positions of the physical
         ECAL layers and the dashed lines show the definition of the virtual pseudo-layers. 
          a) The $xy$-view showing the \CALICE\ stave structure for the ECAL.
              Here hits in the first layer of the stave can be deep in the
             overall calorimeter.  
          b) The $xz$-view showing a possible layout for the ECAL barrel/endcap overlap
             region. Here the pseudo-layers are defined using the projection back to 
             the IP such that the pseudo-layer is closely related
             to the depth in the calorimeter.}
\label{fig:pseudolayer}
\end{figure}

The \PANDORAPFA\ reconstruction is designed to minimise the dependence on the 
detector geometry to enable comparisons of different detector designs.  For this
reason, information is added to the digitised calorimeter hits such that they
become self describing. For example, the {\tt ExtendedCaloHit} objects store 
the size of the corresponding detector pixel. To reduce the dependency 
of the clustering algorithms on
the detector geometry,  hits are ordered in increasing depth in
the calorimeter. This is achieved by defining ``pseudo-layers'' which follow the 
general layer structure of the calorimeters.
This is necessary for calorimeter layouts such as  
in the ECAL stave-like structure being studied by the \CALICE\ collaboration,
shown schematically in Figure~\ref{fig:pseudolayer}. Here there are regions where the 
first layer in a calorimeter stave can be deep in the overall calorimeter structure.

\subsubsection{Isolation Requirements}

Low energy neutrons produced in hadronic showers can travel a significant
distance from the point of production and thus produce isolated energy deposits.
For PFlow calorimetry, these energy deposits are of little use as it is 
impossible to unambiguously associate them with a particular hadronic shower.
For this reason, and to improve the performance of the clustering algorithms,
isolated hits are identified and excluded from the initial cluster finding.
Isolated hits are defined using one of two possible criteria: i) less than a minimum number
of calorimeter hits within a pre-defined distance from the hit in question; or
ii) a cut on the local weighted hit number density, $\rho_i$, defined by:
\begin{eqnarray*}   
                \rho_i   =  \sum_j  w_{ij} &=& \sum_j \frac{1}{(r^\perp_{ij})^n} \\
  \mathrm{where} \ \ \ \ \ \  r^\perp_{ij} &=&  \frac{{\bf r_i  \times} ({\bf r_i} - {\bf r_j} )}{|{\bf r_i}|}. 
\end{eqnarray*}   
Here $\bf{r_i}$ is the position of the hit in question,  the 
sum over $j$ is for all hits within
a certain number of pseudo-layers of hit $i$, and the default value for $n$ is 2. 
By default, method i) is used.

\subsubsection{MIP Identification}
Hits which are consistent with having originated
 from a minimum ionising particle (MIP) are flagged based on
energy deposition and the surrounding hits in the same calorimeter layer. For a hit
to be tagged as MIP-like: a) the energy deposition must be no more than 
{\tt MipLikeMipCut [5.0]} times the mean expected MIP signal, and b) of the adjacent 
(usually 8) pixels in the same layer, no more than {\tt MipMaxCellsHit [1]} should have 
hits above threshold. 
This
information is used in the identification of minimum ionising tracks within
the calorimeter.

\subsubsection{Hit Ordering}

Prior to applying the clustering algorithm, hits within each pseudo-layer are 
ordered either by energy (the default) or by local hit density, $\rho_i$,
 defined above.
The latter option is intended primarily to be used for the case of 
digital calorimetry, where a simple hit count replaces the
analogue energy information.

\subsection{Clustering} 

\label{sec:clustering}

The main clustering algorithm of \PANDORAPFA\ is a cone-based forward projective 
method working from innermost to outermost pseudo-layer. In this manner hits are either
added to existing clusters or they are used to seed new clusters. Throughout the algorithm clusters 
are assigned a direction (or potentially directions) in which they are propagating. 
This allows the clustering algorithm to follow tracks in the calorimeters.
The input to the clustering algorithm is a vector of hits ({\tt ExtendedCaloHit}s) ordered by 
pseudo-layer and energy (or local hit density) and a vector of tracks ({\tt ExtendedTrack}s).

The algorithm starts by {\em seeding} clusters
using the projections of reconstructed tracks onto the front face of the ECAL. 
The initial direction of a track-seeded cluster is obtained 
from the track direction at the ECAL front face. 
The hits in each subsequent pseudo-layer are 
then looped over. Each hit, $i$, is compared to each clustered hit, $j$, 
in the previous layer. The vector displacement, ${\bf r_{ij}}$, is 
used to calculate the parallel and perpendicular displacement of the 
hit with respect to the unit vector(s) ${\bf\hat{u}}$ describing the
cluster propagation 
direction(s), $d_\parallel = {\bf r_{ij} .\hat{u}}$  and 
$d_\perp = |{\bf{r_{ij}\times\hat{u}}}|$. 
Associations are made using a cone-cut, 
$d_\perp < d_\parallel\tan A + b D_{\mathrm{pad}}$, 
where $A$ is the cone 
half-angle, $D_{\mathrm{pad}}$ is the size of a sensor pixel in the layer being 
considered, and $b$ is the number of pixels added to the cone radius. 
Different values of $A$ 
and $b$ are used for the ECAL and HCAL with the default values set to
$\{\tan A_{\mathrm{E}} = 0.3, b_{\mathrm{E}} =1.5\}$ 
and $\{\tan A_{\mathrm{H}} = 0.5, 
b_{\mathrm{H}}=2.5\}$ respectively. The values can be modified
using the steering parameters  
{\tt ClusterFormationAngle} and {\tt ClusterFormationPads}.
For hits in layer $k$, associations are first searched for in layer $k-1$. 
If no association is made, possible associations with clustered hits in 
layers $k-2$ and $k-3$ are considered in turn. If still no association is made, 
associations can be made with nearby hits in existing clusters in the same
pseudo-layer as the hit in question, providing the distance between the
hit centres is less than {\tt SameLayerPadCut = [2.8] ([1.8])} for 
pixels in the ECAL (HCAL). 
If a hit remains unassociated, it is used to 
seed a new cluster. Clusters seeded with calorimeter hits are assigned an
initial direction corresponding to radial propagation from the IP.
This procedure is repeated sequentially for the hits in each 
pseudo-layer working outward from ECAL front-face.    
 
\subsubsection{Fast Photon Identification}

Clusters which are consistent with being from EM showers from photons 
are identified. For reasons of speed, simple cut based 
criteria are used. The fast photon identification 
requirements\footnote{The 
exact cut values depend on the cluster energy and the values
below are those given in the text are the default values for a 
10\,GeV cluster.} are: no associated track; the cluster must start
within 10\,$X_0$ of the front face of the ECAL; the cluster direction
(obtained from a 
linear fit to the energy-weighted centroids of the hits in each pseudo-layer)
must point to within $20^\circ$ of the IP; 
the rms deviation of the 
hits in the cluster around the linear fit to the centroids in each
calorimeter layer must be less than 40\,cm; and 
the fraction of hits classified as MIP-like must be less than 30\,\%.
In addition, weak cuts on the longitudinal development of the shower
are imposed. Photon clusters are essentially frozen at this stage in
the \PANDORAPFA\ algorithm; they are not used in the subsequent 
topological cluster merging or reclustering algorithms.

\subsubsection{Photon Clustering (optional)} 
	
Rather than attempting to cluster all calorimeter hits in a single pass, 
\PANDORAPFA\ can be run in a mode ({\tt PhotonClustering} $>$ 0) 
where the clustering algorithm described above 
is first applied solely to the ECAL hits to identify photons
as the first stage of PFlow reconstruction. The clustering algorithm parameters are 
chosen to reflect the narrowness of EM showers.
Reconstructed 
clusters which are consistent with the expected EM transverse and longitudinal 
shower profiles
(see Section~\ref{sec:photonID}) are stored and the associated calorimeter hits
are not considered in the second pass
of the clustering algorithm. The identified photon
clusters are added back to the event 
just prior to the formation of the PFOs. For the results presented
in this paper, photon clustering is run prior to the main clustering algorithm.

\subsection{Topological Cluster Merging} 

\label{sec:topological}

\begin{figure}[hbtp]
\epsfxsize=1.0\columnwidth
\centerline{\includegraphics[width=0.57\columnwidth,angle=270]{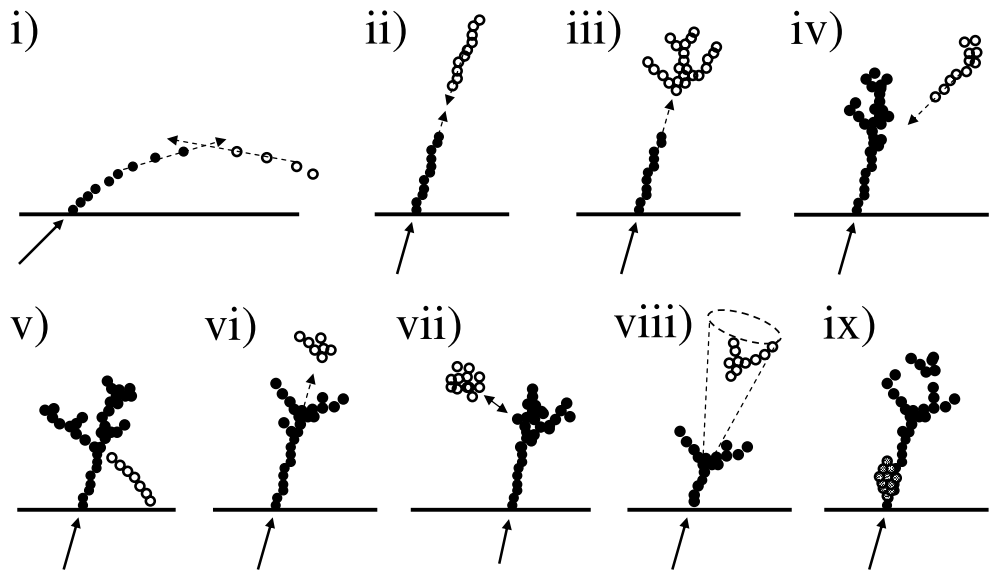}}
\caption{The main topological rules for cluster merging: i) looping track segments;
         ii) track segments with gaps; iii) track segments pointing to 
             hadronic showers; iv) track-like neutral clusters pointing
             back to a hadronic shower; v) back-scattered tracks from
             hadronic showers; vi) neutral clusters which are close to 
             a charged cluster; vii) a neutral cluster near to a charged cluster; 
             viii) cone association; and ix) recovery of photons which overlap
             with a track segment.
             In each case the arrow indicates the track, the filled points
           represent the hits in the associated cluster and the open points
           represent the hits in the neutral cluster.}
\label{fig:topology}
\end{figure}

By design the initial clustering algorithm errs on the side of splitting up 
true clusters rather than merging energy deposits from more than one particle
into a single cluster. Hence, the next stage in the \PANDORAPFA\ algorithm 
is to merge clusters which are not already associated to tracks 
(termed ``neutral clusters'') with clusters which have an associated 
track (termed ``charged clusters''). The merging algorithms are based on the 
clear topological signatures 
shown schematically in Figure~\ref{fig:topology}.

This procedure takes advantage
of the high granularity of the ECAL and HCAL of a detector designed for
PFlow reconstruction.  
For clusters with an associated
track, the location of the first hadronic interaction is identified and
the properties of the track-like segment in the calorimeter 
before the interaction are reconstructed.
For neutral clusters,  track-like segments are identified in the first
and last six pseudo-layers of the cluster based on 
fraction of hits classified as MIP-like and the rms 
deviation of the hit positions about straight line fits. For track-like
segments, the fitted line, $\bf{r_0+\kappa\hat{d}}$,
is used to project forwards or backwards 
in the calorimeter. Similarly, the entire cluster may be classified as track-like. 
The main topological rules for cluster association are:
\begin{enumerate}[(i)]
\item Looping tracks: Because of the forward projective nature of the primary 
       clustering algorithm, tracks which turn back in the calorimeter due to the
       high magnetic field are often reconstructed as two track-like clusters. 
       The track-like segments at the ends of the clusters are projected forwards and the
       clusters are combined if the distance of closest approach of the two forward-going 
       track projections is less than {\tt LooperClosestApproachCutECAL [5\,cm]}. 
\item Broken tracks: Non-continuous tracks in the calorimeters can arise when particles
       cross boundaries between physical sub-detectors or cross dead regions
       of the calorimeters. Such instances are identified 
       using track-like segments in the last six layers of a charged cluster and
       the first six layers in a neutral cluster.  The clusters may be merged if the distance 
       of closest
       approach of the forward-going and backward-going track-like segment projections 
       is less than {\tt TrackMergeCutEcal [2.5\,cm]}.
\item Tracks pointing to showers: If, when projected forward, a track-like charged cluster 
       points to within {\tt TrackMergeCutEcal [2.5\,cm]} of the start of a cluster deeper in
       the calorimeter, the clusters may be merged.
\item Track-like clusters pointing back to hadronic interactions: If the start of a neutral
       cluster is a track-like segment and it points to within
       {\tt TrackBackMergeCut [3.0\,cm]} of the identified first hadronic interaction of
       charged cluster, the clusters may be merged.    
\item Back-scattered tracks: Hadronic interactions can produce tracks in the calorimeter
       which propagate backwards in the calorimeter. Due to the forward projective nature
       of the clustering algorithm, these often will be reconstructed as separate clusters. 
       Back-scattered tracks are identified as track-like clusters which point to within 
       {\tt TrackBackMergeCut [3.0\,cm]} of the identified hadronic interaction of
       a charged cluster. 
\item Hadronic interactions pointing to neutral clusters: If a charged-cluster has track-like
      segment prior to the identified interaction point, and it points to within 
      {\tt TrackForwardMergeCut [5.0\,cm]} 
      of the start of a cluster deeper in the calorimeter, the clusters may be merged.
\item Proximity-based merging:
       The minimum distance between a charged cluster, of energy $E_C$, and a neutral cluster, 
       of energy $E_N$, is defined as the 
       smallest distance between any of the hits in the two clusters. If this
       distance is less than 
       {\tt ProximityCutDistance [5\,cm]} then the clusters
       maybe merged if there is additional evidence that the two 
       clusters originate from a single hadronic shower.
       To suppress false matches the
       $\chi^2$ consistency between the original and merged cluster energies and the
       associated track momentum, $p$, is used. 
       The merged cluster energy, 
       $E^\prime=E_C+E_N$, must be consistent with the track momentum,   
       $\chi^\prime = (E^\prime-p)/\sigma_{E^\prime}<${\tt EnergyChi2ForCrudeMerging [2.5]},
       where $\sigma_{E^\prime}$ is the uncertainty on the merged cluster energy assuming
       that it is a hadronic shower. In addition, the $\chi^2$ consistency must not be 
       significantly worse than
       that for the original cluster, $\Delta\chi^2 = (\chi^{\prime})^2 - \chi^2 < $
       {\tt EnergyDeltaChi2ForCrudeMerging [1.0]}, where $\chi = (E_C - p)/\sigma_{E_C}$.
\item Cone-base merging: Starting from the identified hadronic interaction point of 
       each charged cluster, a cone of half-angle {\tt CosineConeAngle [0.9]} is defined
       in the direction of the track-like segment of the cluster. Neutral clusters 
       deeper in the calorimeter with more than 50\,\% of the energy of lying within this 
       cone may be merged providing the above $\chi^2$ consistency requirements are
       satisfied. If there is no track-like segment at the start of the charged cluster,
       the track direction is used.
\item Photon recovery: In dense jets minimum ionising particles may pass through 
        the EM shower from a photon, resulting in a single reconstructed
        cluster. Cases where the hadron interacts a significant distance after the
        end of the EM shower are identified and photons overlapping with
        charged clusters are recovered. 
\end{enumerate}

\subsection{Re-clustering} 

\begin{figure}[htb]
\epsfxsize=1.0\columnwidth
\centerline{\includegraphics[width=1.0\columnwidth,angle=270]{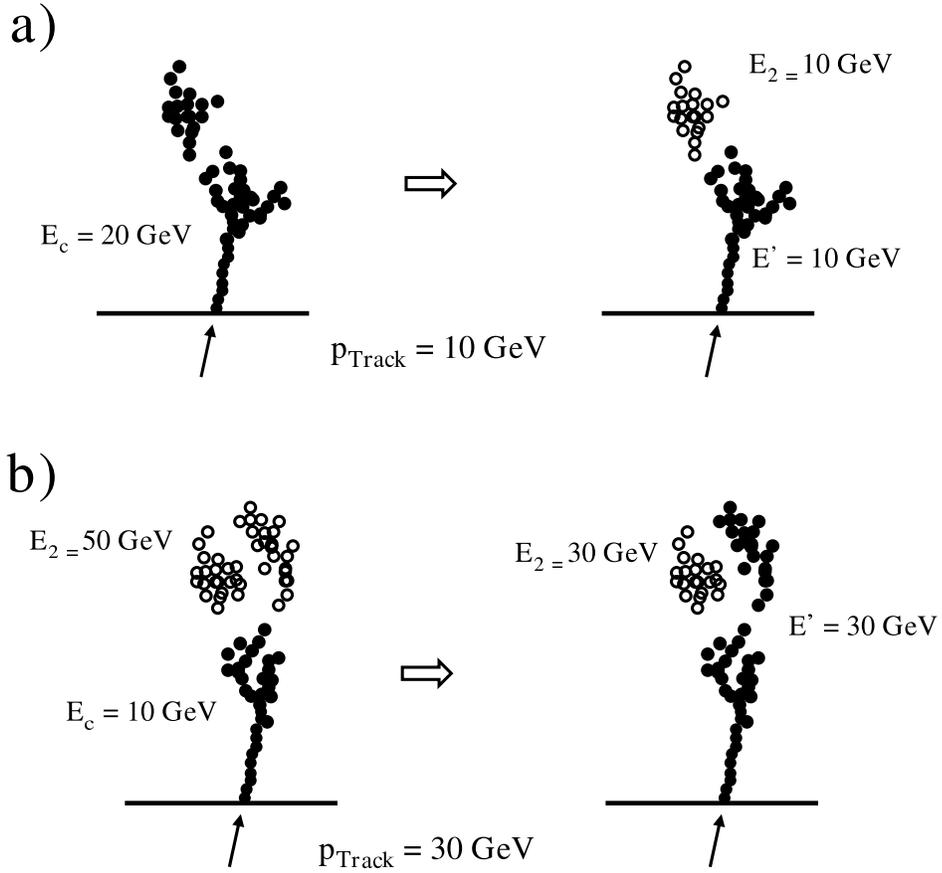}}
\caption{Schematic examples of the main reclustering strategies used in \PANDORAPFA.
         The arrows indicates the track, the filled points
         represent the hits in the associated charged cluster and the open points
         represent the hits in the neutral cluster.
         a) Here the charged cluster energy is initially significantly greater than the associated
            track momentum. The hits are reclustered using modified parameters for the
            clustering algorithm in the hope that a more consistent solution can be found.
         b) Here the cluster energy is significantly less than the associated track momentum.
            The topological association algorithms vii) and viii) have not added the
            neutral cluster as his would have resulted in a charged cluster with too much
            energy for the track momentum. The hits are reclustered in the hope that the 
            neutral cluster naturally splits in
            such a way that the topological association algorithm will now make the correct
            association.
\label{fig:reclustering}}
\end{figure}

The previous four stages of the \PANDORAPFA\ algorithm are found to 
perform well for jets with energy less than about $50$\,GeV. 
At higher energies the jet energy resolution 
degrades due to
the increasing overlap between the hadronic showers from different 
particles. It is possible to detect
such reconstruction failures by comparing the charged cluster
energy, $E_C$, with the momentum of the associated track, $p$. 
A possible reconstruction failure is identified if 
$|(E_C-p)/\sigma_{E_C}| >$ {\tt ChiToAttemptReclustering [3.0]}.
In this case the \PANDORAPFA\ algorithm attempts to find a
more self-consistent clustering of the calorimeter hits.
If, for example, 
a 10\,GeV track is associated with a 20\,GeV 
calorimeter cluster, 
shown schematically in Figure~\ref{fig:reclustering}a),
a potential reconstruction failure is identified. 
One possible
approach would be to simply remove hits from the cluster until
the cluster energy matched the track momentum. However, this
does not use the full information in the event. Instead, the
clustering algorithm is modified  iteratively with the hope that 
a more correct clustering of the hits will be found. This is implemented 
by passing the hits in the cluster and the 
associated track(s) to the main clustering algorithm 
described in Sections~\ref{sec:clustering} and 
\ref{sec:topological}.  The algorithm is applied repeatedly, 
using successively smaller values  of the parameters $A$ 
and $b$, with the aim of splitting the original cluster 
so that the track momentum and associated cluster energy are
compatible, as
indicated in Figure~\ref{fig:reclustering}a). 
In principle, completely different clustering algorithms could be tried.  
In cases where no
significant improvement in the $\chi^2$ compatibility of the 
track and associated cluster is found, the original cluster 
is retained. 

In steps vii) and viii) of the topological clustering, described in
Section~\ref{sec:topological}, the case where too little energy is
associated with the track is addressed. However, in a dense jet environment,
the neutral cluster which should be associated with a charged cluster
may itself be merged with
another neutral cluster, as indicated in Figure~\ref{fig:reclustering}b).
In such cases the reclustering procedure acts on the combination of
hits in the charged cluster associated to the track and 
nearby neutral clusters.

\subsection{Photon Identification and Recovery}

\label{sec:photonID}
 
A relatively sophisticated photon identification algorithm is applied to the 
reconstructed clusters.
The longitudinal profile of the energy deposition, $\Delta E_{obs}$, as a function of number of
radiation lengths from the shower start, $t$, is compared to that expected~\cite{bib:pdgem} 
for an EM shower:
\begin{eqnarray*}
 \Delta E_{EM}&\approx&E_0 \frac{(t/2)^{a-1} e^{-t/2}} {\Gamma(a)}  t, \\  
   \ \ \ \mathrm{where} \ \ \ a&=&1.25 +\frac{1}{2}\ln\frac{E_0}{E_c},
\end{eqnarray*} 
$E_0$ is the shower energy and $E_c$ is the critical energy, which is chosen to give the
appropriate average MC shower profile in the ECAL. The level of agreement is
parameterised by the sum over samplings in radiation length of the fractional deviation
of the cluster profile compared the expectation for an EM shower:
\begin{eqnarray*} 
   \delta &=& \frac{1}{E_0} \sum_{i} | \Delta E_{obs}^i - \Delta E_{EM}^i |.    
\end{eqnarray*}
This approach was preferred to a $\chi^2$-based metric  as it is less sensitive
to large local deviations which might arise from energy deposits from other nearby
particles. The quantity $\delta$ is minimised as a function of the assumed starting
point of the shower, $t_0$. Hence the output of the shower shape algorithm is a measure of
the consistency with the expected EM shower profile, $\delta$, and the
starting depth of the shower in the ECAL, $t_0$ (in radiation lengths). These variables are
used as the basis for identifying clusters as photons. Transverse information is not used
as this would make the photon identification algorithm more sensitive to over-lapping 
EM showers from very close photons.

\subsubsection{Photon Recovery}

The compact nature of EM showers is utilised in an attempt to identify photons 
which may have been merged into the cluster associated with a hadronic shower. 
The transverse energy distribution (ECAL only) of the reconstructed clusters is 
determined assuming
that the cluster originates from the IP. A peak finding algorithm attempts
to identify localised energy depositions which are displaced from the associated
track. If the longitudinal energy profile in these regions is consistent with
being an EM shower, the relevant hits are removed from the cluster and
used to form a new cluster (assumed to be a photon). Cases where removing the 
candidate photon would result in the remaining cluster energy being inconsistent
with the associated track momentum are vetoed.

\subsection{Fragment Removal} 
 At this late stage in \PANDORAPFA\ there are still a significant number of  
 ``neutral clusters'' (not identified as photons) which are {\em fragments} 
       of charged particle hadronic showers. An attempt is made to
      identify these clusters and merge them with the appropriate parent charged cluster. 
      All non-photon neutral clusters, $i$, are compared to all charged clusters, $j$.
      For each pair of clusters a quantity, $e_{ij}$, is defined which encapsulates the evidence 
      that cluster $i$ is a fragment from cluster $j$ using the following information:
           the number of calorimeter layers in which the minimum distance between 
           the hits in the two clusters are separated by less than 
           {\tt FragmentRemovalContactCut [2]} pixels; the fractions of the energy of cluster $i$
           within three narrow cones defined by the first hadronic interaction in cluster $j$;
           the minimum distance of the centroid within a layer of cluster $i$ to the fitted
           helix describing the track associated to cluster $j$; and the minimum distance between any of the
           hits in the two clusters.  
      The requirement, $R_{ij}$, for the 
      clusters to be merged,
      {\em i.e.} the cut on $e_{ij}$, depends on the location of the depth of the neutral cluster 
      in the calorimeter and
      the change in the  $\chi^2$ for the track$-$cluster energy consistency that would occur
      if the clusters were merged, 
        $$\Delta\chi^2 = (p-E_j)^2/\sigma^2_{E_j} - (p-E_j-E_i)^2/\sigma^2_{E_{ij}}.$$ 
      If $e_{ij} > R_{ij}$ the clusters are
      merged. This {\em ad hoc} procedure gives extra weight to cases 
      where the consistency of the track momentum and associated cluster 
      energy improves
      as a result of merging the neutral cluster with the charged cluster.

\subsection{Formation of Particle Flow Objects}
 
\begin{figure}[hbtp]
\epsfxsize=1.0\columnwidth
\centerline{\includegraphics[width=1.00\columnwidth]{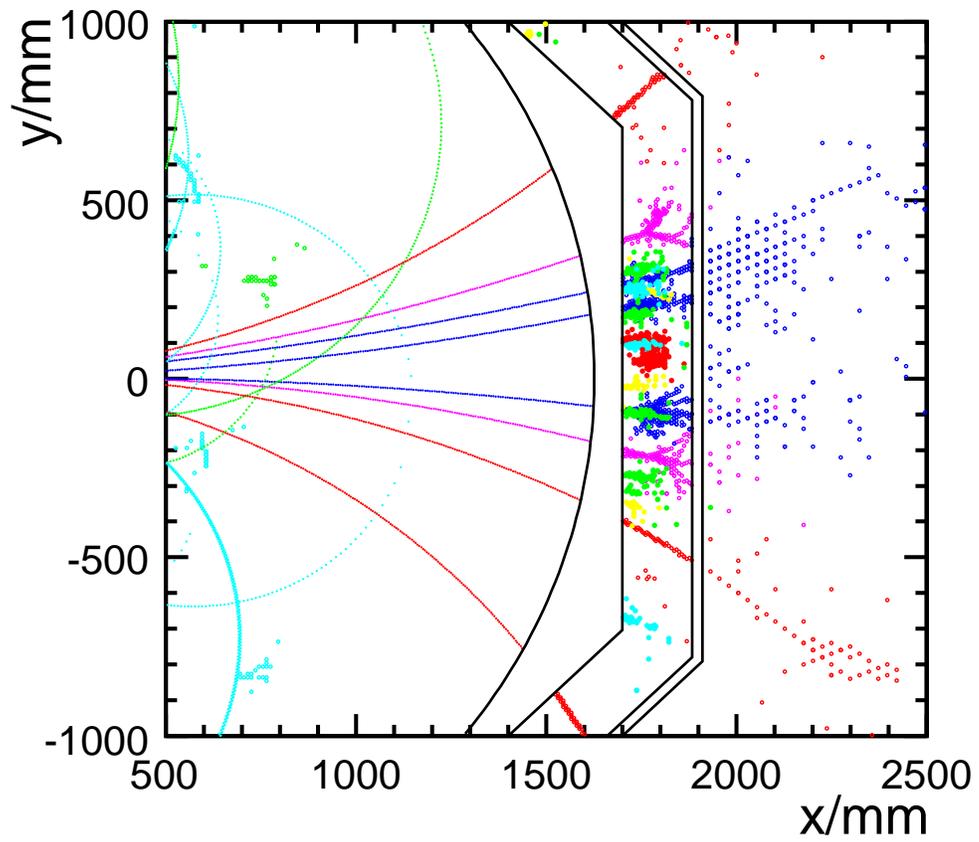}}
\caption{ \PANDORAPFA\ reconstruction of a 100\,GeV jet in the \MOKKA\
             simulation of the ILD detector. The different PFOs are shown 
             by colour/grey-shade according to energy.}
\label{fig:jet100}
\end{figure}

     The final stage of \PANDORAPFA\ is to create Particle Flow Objects (PFOs) from
     the results of the clustering. Tracks are matched to clusters on the basis 
     of the distance closest approach of the track projection into the first 10 layers 
     of the calorimeter. If a hit is found within a distance {\tt TrackClusterAssociationDistance [10\,mm]} of the track extrapolation, an 
     association is made. 
     If an identified kink is consistent with being from a $K^\pm\rightarrow\mu^\pm\nu$ or
     $\pi^\pm\rightarrow\mu^\pm\nu$ decay the parent track is used to form the PFO, otherwise
     the daughter track is used.  
     Relatively primitive particle identification is applied and the 
     reconstructed PFOs, including four-momenta, are written out in \LCIO~\citep{bib:lcio} 
     format. Figure~\ref{fig:jet100}a) shows an example of a \PANDORAPFA\ reconstruction of 
     a 100\,GeV jet from a $\Zzero\rightarrow\uu$ decay at $\roots=200$\,GeV. The ability
     to track particles in the high granularity calorimeter in the ILD detector
     concept can be seen clearly.

\section{\bf\boldmath Parameterising Particle Flow Performance: \rms}

\begin{figure}[hbtp]
\epsfxsize=1.0\columnwidth
\centerline{\includegraphics[width=1.00\columnwidth]{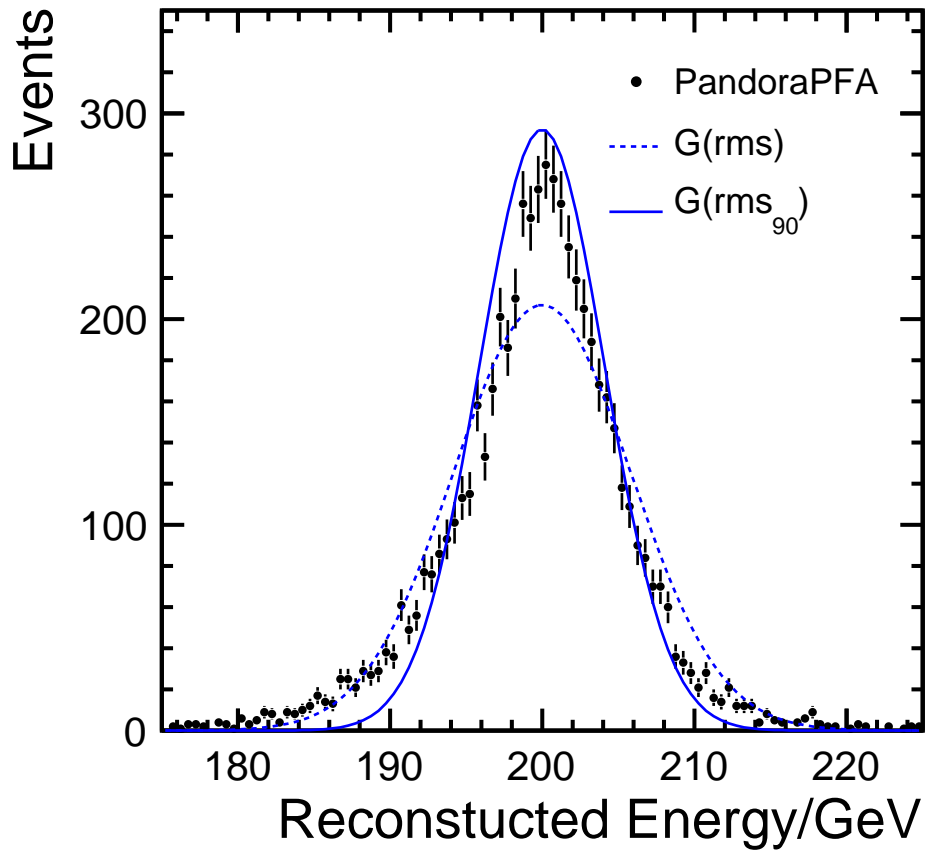}}
\caption{ The total reconstructed energy from reconstructed 
             PFOs in 200\,GeV $\Zzero\rightarrow{\mathrm{uds}}$ events for initial quark directions
             within the polar angle acceptance $|\cosqq|<0.7$. 
             The dotted line shows the best fit Gaussian distribution
             with an rms of 5.8\,\GeV. The solid line shows
             a Gaussian distribution, normalised to the same number of events,
             with standard deviation equal to $\rms$ ({\it i.e. }$\sigma =4.1$\,GeV).}
\label{fig:jet100res}
\end{figure}

Figure~\ref{fig:jet100res} shows the distribution of PFA reconstructed 
energy for simulated  
$(\Zzero/\gamma)^*\rightarrow\qq$ events (light quarks only, {\it i.e.} q=u,d,s)
generated at $\roots=200$\,GeV with the $\Zzero$ decaying at rest,
termed ``$\Zzero\rightarrow{\mathrm{uds}}$'' events.
A cut on the polar angle of the generated $\qq$ system, $\thetaqq$,
is chosen to avoid the barrel/endcap overlap region,  $|\cosqq|<0.7$. Only
light quark decays are considered as, currently, \PANDORAPFA\ does not include
specific reconstruction algorithms to attempt to recover missing energy from semi-leptonic
decays of heavy quarks. 
The reconstructed energy distribution of Figure~\ref{fig:jet100res}
is not Gaussian. This is not surprising;  one might expect a Gaussian core for 
perfectly reconstructed events, and tails corresponding to the
population of events where confusion is significant.
Quoting the rms, in this case 5.8\,GeV, as a measure of the jet 
energy resolution over-emphasises 
the importance of these tails. 
In this paper, performance is quoted in 
terms of $\rms$, which is defined as the rms in the smallest range of reconstructed 
energy which contains 90\,\% of the events. 
For the data shown in Figure~\ref{fig:jet100res},
$\rms = 4.1$\,GeV (equivalent to a single jet energy resolution 
of 2.9\,\%). The advantage of using $\rms$ is that it is robust and is relatively
insensitive to the tails of the distribution; it parameterises the 
resolution for the bulk of the data. One possible criticism of this 
performance measure is that 
for a true Gaussian distribution, 
$\rms$ would be 21\,\% smaller than the true rms.
However, for the non-Gaussian distribution from PFlow reconstruction, 
this is not a fair comparison. For example, the central region
of the reconstructed energy distribution\footnote{Here the best 
fit Gaussian to the region $196-205$\,GeV has 
an rms of 3.5\,\GeV} is $15\,\%$ {\em narrower} than 
the equivalent Gaussian of  $\sigma = \rms$ as shown in Figure~\ref{fig:jet100res}.
To determine the equivalent Gaussian statistical 
power, a MC study was performed assuming a signal with the shape
of the PFA reconstructed energy distribution centred on $x$ and a flat background. 
A fit to determine the value of $x$ was performed using the shape of the 
PFA distribution as a resolution function (fitting template).
The process was repeated assuming a signal with same number of events but now with
a Gaussian energy distribution. The width of the Gaussian (for both 
the signal and the fitting function)
was chosen to give the same statistical precision on $x$ as obtained with the PFA resolution
function. From a fit to signal and background components the same fitted uncertainty,
$\sigma_x$, is obtained for
a Gaussian with standard deviation of $1.1\times\rms$.
On this basis it is concluded that the statistical power for PFlow reconstruction
with \PANDORAPFA\ yielding $\rms$ is equivalent to a 
Gaussian resolution with $\sigma = 1.1\times\rms$.  This conclusion
does not depend strongly on the assumed relative normalisation of the signal and 
background or the total energy of the generated events.

\section{Particle Flow Performance}

\begin{figure}[htb]
\epsfxsize=1.0\columnwidth
\centerline{\includegraphics[width=1.0\columnwidth]{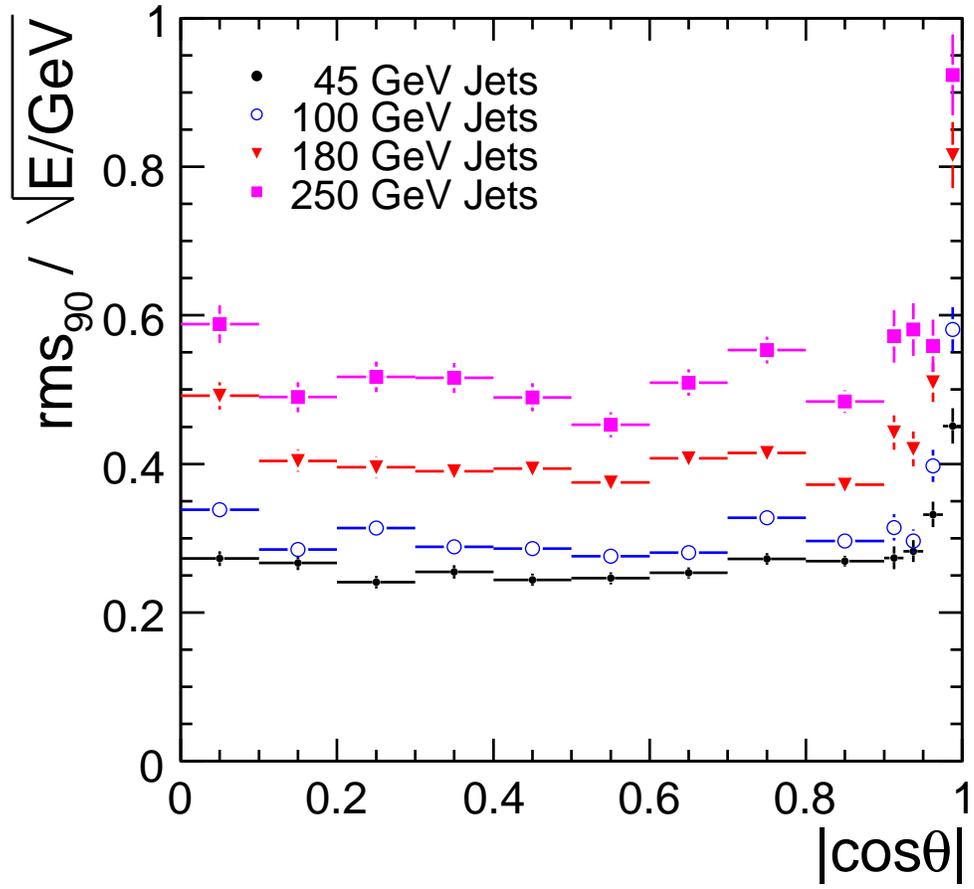}}
\caption{The jet energy resolution, defined as the $\alpha$ in $\sigma_E/E=\alpha/\sqrt{\EGeV}$, 
        plotted versus $\cosqq$ for four different values of $\roots$. The plot
        shows the resolution obtained from  $(\Zzero/\gamma)^*\rightarrow\qq$ events (q=u,d,s)
        generated at rest. 
    \label{fig:rmsVersusTheta}}
\end{figure}

The performance of the \PANDORAPFA\ algorithm with the ILD 
detector concept is studied using MC samples of approximately 10000
$\Zzero\rightarrow{\mathrm{uds}}$ generated with the $\Zzero$ decaying 
at rest with 
$E_Z$ = 91.2, 200, 360, and 500\,GeV. These jet energies are typical
of those expected at the ILC operating at $\roots = 0.5-1.0$\,TeV. In addition, to
study the performance at higher energies, events were generated with $E_Z=750$\,GeV and 1\,TeV. 
Jet fragmentation and hadronisation was performed using the \PYTHIA~\citep{bib:pythia}
program tuned to the fragmentation data from the \OPAL\ experiment~\citep{bib:opalpythia}.  
The events were passed through the \MOKKA\ simulation of the ILD
detector concept which is described in detail in~\cite{bib:ildloi}. 
The {\tt LCPHYS}\cite{bib:lcphys} \GEANT\ physics list was used for the modelling of
hadronic showers. For each set of events, the total energy is reconstructed and the
jet energy resolution is obtained by dividing the total energy resolution
by $\sqrt{2}$. 
Figure~\ref{fig:rmsVersusTheta} shows the jet energy resolution 
as a function of the polar angle of the quarks in $\Zzero\rightarrow\qq$ events.
The energy resolution does not vary significantly in the region 
$|\cos\theta|<0.975$. A small degradation in the energy resolution is 
seen for the barrel-endcap overlap 
region, $0.7<|\cos\theta|<0.8$. In addition, there is a small
degradation in performance at $\cos\theta\approx 0$ due to the 
TPC central membrane and 
gaps between sections of the HCAL as simulated in the ILD detector model.

\begin{table*}[bth]
\begin{center}
\begin{tabular}{|r|rrcc|}
\hline
  {\bf Jet Energy}        & {\bf rms}        & {\bf\boldmath $\rms(\Ejj)$}     &  {\bf\boldmath $\rms(\Ejj)/\sqrt{\Ejj}$} & {\bf\boldmath $\rms(\Ej)/\Ej$ } \\ \hline
  45 GeV            &  3.4\,GeV  &  2.4\,GeV  &  25.2\,\%   &   $(3.74\pm0.05)\,\%$  \\
  100 GeV           &  5.8\,GeV  &  4.1\,GeV  &  29.2\,\%   &   $(2.92\pm0.04)\,\%$  \\
  180 GeV           & 11.6\,GeV  &  7.6\,GeV  &  40.3\,\%   &   $(3.00\pm0.04)\,\%$  \\
  250 GeV           & 16.4\,GeV  & 11.0\,GeV  &  49.3\,\%   &   $(3.11\pm0.05)\,\%$  \\ \hline
  375 GeV           & 29.1\,GeV  & 19.2\,GeV  &  81.4\,\%   &   $(3.64\pm0.05)\,\%$  \\
  500 GeV           & 43.3\,GeV  & 28.6\,GeV  &  91.6\,\%   &   $(4.09\pm0.07)\,\%$  \\ \hline
\end{tabular}
\caption{Jet energy resolution for $\Zzero\rightarrow$uds events with $|\cosqq|<0.7$, 
            expressed as: i) the rms of the reconstructed
            di-jet energy distribution, $\Ejj$; ii) $\rms$ for $\Ejj$;  iii) the effective
            constant $\alpha$ in $\rms(\Ejj)/\Ejj = \alpha(\Ejj)/\sqrt{\EjjGeV}$; and iv) the
            fractional jet energy resolution for a single jet where $\rms(\Ej)=\rms(\Ejj)/\sqrt{2}$.
\label{tab:resvsE}}
\end{center}
\end{table*}

Table~\ref{tab:resvsE} summarises the current performance of the 
\PANDORAPFA\ algorithm applied to ILD detector simulation.
For the typical ILC jet energy range, $45-250$\,GeV,  
the energy resolution is significantly better 
than the best resolution achieved 
at LEP, $\sigma_E/E \approx 0.65/\sqrt{E(\GeV)}$.
Table~\ref{tab:resvsE} also lists the single jet energy resolution. 
For jet energies in the range $45-375$\,GeV this is better than 3.8\,\%, 
which is necessary to resolve hadronic decays of $\Wboson$ and $\Zzero$ bosons.
These results clearly demonstrate the potential of PFlow calorimetry
at the ILC; the jet energy resolution obtained is approximately
a factor two better than might be achievable with a traditional 
calorimetric approach. Furthermore, it is expected that the performance
of \PANDORAPFA\ will improve with future refinements to the algorithm. 

It is worth noting, that for perfect PFlow reconstruction, the 
energy resolution would be described by $\sigma_E/E \approx \alpha/\sqrt{\EGeV}$,
where $\alpha$ is a constant. The fact that  
this does not apply is not surprising; 
as the particle density increases it becomes harder to correctly
associate the calorimetric energy deposits to the particles and the 
confusion term increases. Also it should be noted that in a physics
analysis involving multi-jet final states, the resolution may be
degraded by imperfect jet finding.

\section{Understanding Particle Flow Performance}

\label{sec:perf}

\PANDORAPFA\ is a fairly complex algorithm, consisting of over 10,000 lines of C++. It has 
a number of distinct stages which interact with each other in the sense that reconstruction 
failures in one part of the software can be corrected at a later stage. 
The relative importance of the different stages in the reconstruction is investigated 
by turning off parts of the \PANDORAPFA\ algorithm.
Table~\ref{tab:recodep} compares the full \PANDORAPFA\ reconstruction with the 
algorithm run: 
a) without the topological cluster merging phase;
b) without the reclustering phase; 
c) without running the photon clustering stage prior to the running the full clustering; 
d) without fragment removal; and 
e) the case where tracks from $V^0$s and kinks are not used in the event reconstruction.
There are a number of notable features. The topological clustering and fragment removal algorithms
are important at all energies. For low energy jets, the reclustering stage is not particularly important.
This is because the primary clustering and topological clustering algorithms are sufficient 
in the relatively low particle density environment. With increasing jet energy,
the reclustering stage becomes more important. For high energy jets ($E>100$\,GeV) 
it is the single most important step in the reconstruction after the initial clustering. 
Running the dedicated photon clustering stage before the main clustering algorithm is 
advantageous for higher energy jets.
The $V^0$/kink finding does not significantly improve the resolution, although it is an important
part in the identification of the final reconstructed particles. 

\begin{table*}[bth]
\begin{center}
\begin{tabular}{|l|cccc|}
\hline
 {\bf Algorithm}             & \multicolumn{4}{c|}{\bf \boldmath Jet Energy Resolution $\rms(\Ej)/\Ej$ [\%] }                          \\  
                             & $\Ej$=45\,GeV        & $\Ej$=100\,GeV      & $\Ej$=180\,GeV      &   $\Ej$=250\,GeV    \\ \hline
Full \PANDORAPFA\            & $3.74\pm0.05$  & $2.92\pm0.04$ & $3.00\pm0.04$ & $3.11\pm0.05$ \\ \hline
a) No Topological Clustering & $4.02\pm0.05$  & $3.25\pm0.04$ & $3.52\pm0.05$ & $3.67\pm0.06$ \\
b) No Reclustering           & $3.83\pm0.05$  & $3.30\pm0.04$ & $3.91\pm0.05$ & $4.19\pm0.07$ \\
c) No Photon Clustering Stage& $3.66\pm0.05$  & $2.99\pm0.04$ & $3.13\pm0.04$ & $3.31\pm0.05$ \\
d) No Fragment Removal       & $4.05\pm0.05$  & $3.21\pm0.04$ & $3.25\pm0.04$ & $3.40\pm0.06$ \\
e) No $V^0$/Kink Tracks      & $3.78\pm0.05$  & $2.96\pm0.04$ & $3.02\pm0.04$ & $3.13\pm0.05$ \\ \hline
\end{tabular}
\caption{
Jet energy resolutions ($\rms/E$) for the full \PANDORAPFA\ reconstruction compared 
to that obtained:
a) without the topological cluster merging phase;
b) without the reclustering phase; 
c) without running the photon clustering stage prior to the running the full clustering; 
d) without fragment removal; and 
e) the case where tracks from $V^0$s and kinks are not used in the event reconstruction.
\label{tab:recodep}}
\end{center}
\end{table*}

\begin{figure}[htb]
\epsfxsize=1.0\columnwidth
\centerline{\includegraphics[width=1.0\columnwidth]{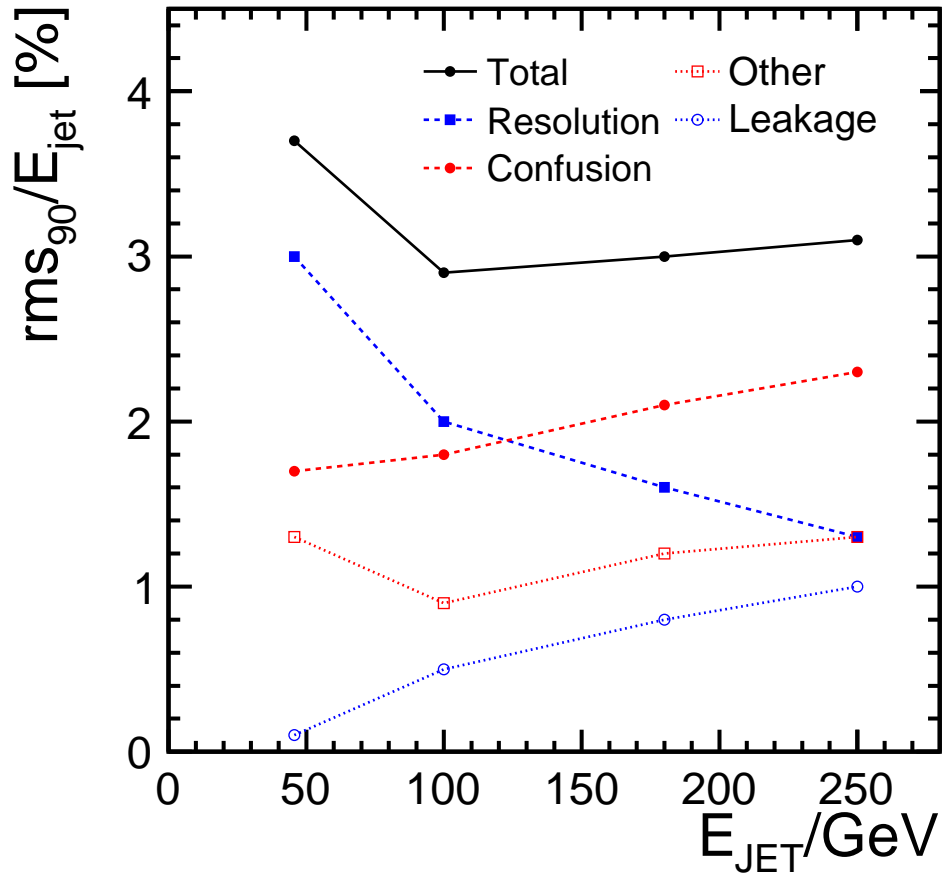}}
\caption{The contributions to the PFlow jet energy resolution obtained with \PANDORAPFA\ as a
          function of energy. The total is (approximately) the quadrature sum of the 
         components.\label{fig:pfacont}}
\end{figure}

The contributions to the jet energy resolution
have been estimated by replacing different steps in \PANDORAPFA\ with
algorithms which use MC information to perform: a) perfect reconstruction 
of photons as the first phase of the algorithm; b) perfect reconstruction of neutral hadrons; 
and c) perfect identification of fragments from charged hadrons. The
jet energy resolutions obtained using these ``perfect'' algorithms enable the contributions
from {\it confusion} to be estimated. In addition, studies using a deep
HCAL enable the contribution from leakage to be estimated. Finally, MC information
can be used to perform ideal track pattern recognition enabling the impact of imperfect track
finding code to be assessed. 
Table~\ref{tab:pfacont} lists the estimated breakdown of the
total jet energy into its components, including the contributions from 
calorimetric energy resolution ({\it i.e.} the energy resolution for photons and neutral 
hadrons). For the current \PANDORAPFA\ algorithm, the contribution from 
the calorimetric energy resolution,
$\approx 21\,\%/\sqrt{E}$, dominates the jet energy resolution for 45\,GeV jets. 
For higher energy jets, the confusion term dominates. This behaviour is summarised
in Figure~\ref{fig:pfacont}. The contributions from resolution and confusion are
roughly equal for 120\,GeV jets. From Table~\ref{tab:pfacont} it can be seen that
the most important contribution for high energy jets is confusion due to neutral
hadrons being lost within charged hadron showers. For all jet energies considered,
fragments from charged hadrons, which tend to be relatively low in energy, 
do not contribute significantly to the jet energy resolution.

\begin{table*}[ht]
\begin{center}
\begin{tabular}{|l|cccc|}
\hline
{\bf Contribution }        & \multicolumn{4}{c|}{\bf\boldmath Jet Energy Resolution $\rms(\Ej)/\Ej$} \\  
                           & $\Ej$=45\,GeV & $\Ej$=100\,GeV  & $\Ej$=180\,GeV & $\Ej$=250\,GeV  \\ \hline
Total                      &     3.7\,\%   &     2.9\,\%     &    3.0\,\%     &  3.1\,\%  \\ \hline
Resolution                 &     3.0\,\%   &     2.0\,\%     &    1.6\,\%     &  1.3\,\%  \\ 
Tracking                   &     1.2\,\%   &     0.7\,\%     &    0.8\,\%     &  0.8\,\%  \\
Leakage                    &     0.1\,\%   &     0.5\,\%     &    0.8\,\%     &  1.0\,\%  \\
Other                      &     0.6\,\%   &     0.5\,\%     &    0.9\,\%     &  1.0\,\%  \\ 
Confusion                  &     1.7\,\%   &     1.8\,\%     &    2.1\,\%     &  2.3\,\%  \\  \hline
i) Confusion (photons)        &     0.8\,\%   &     1.0\,\%     &    1.1\,\%     &  1.3\,\%  \\
ii) Confusion (neutral hadrons)&     0.9\,\%   &     1.3\,\%     &    1.7\,\%     &  1.8\,\%  \\
iii) Confusion (charged hadrons)&     1.2\,\%   &     0.7\,\%     &    0.5\,\%     &  0.2\,\%  \\ \hline
\end{tabular}
\caption{The PFlow jet energy resolution obtained with \PANDORAPFA\ broken down into
          contributions from: intrinsic calorimeter resolution, imperfect tracking, leakage 
          and confusion. The different confusion terms correspond to: i) hits from photons which are lost
          in charged hadrons; ii) hits from neutral hadrons that are lost in charged hadron
          clusters; and iii) hits from charged hadrons that are reconstructed as a neutral
          hadron cluster.
\label{tab:pfacont}}
\end{center}
\end{table*}

\begin{figure}[tbh]
\epsfxsize=1.0\columnwidth
\centerline{\includegraphics[width=1.0\columnwidth]{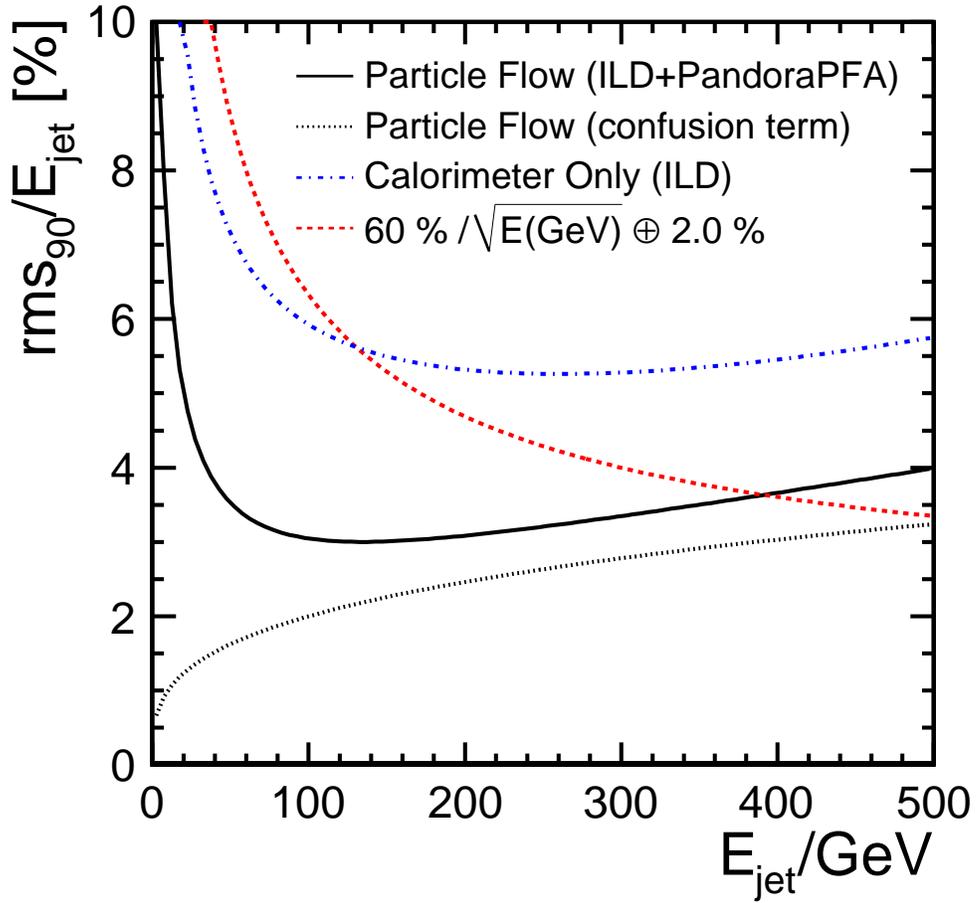}}
\caption{The empirical functional form of the jet energy resolution obtained from PFlow 
          calorimetry (\PANDORAPFA\ and the ILD concept). The estimated
         contribution from the confusion term only is shown (dotted). The dot-dashed curve
         shows a parameterisation of the jet energy resolution obtained from          the total calorimetric energy deposition in the ILD detector. In addition,
         the dashed curve,  
         ${60\,\%}/{\sqrt{E(\mathrm{GeV})}} \oplus 2.0\,\%$, is shown to give an {\rm indication} 
         of the resolution achievable using a traditional calorimetric approach.
        \label{fig:pflow}}
\end{figure}

The numbers in Table~\ref{tab:pfacont} can be used to obtain an semi-empirical
parameterisation of the jet energy resolution:
\begin{eqnarray*}
     \frac{\rms}{E} & = & \frac{21}{\sqrt{E}} 
           \oplus 0.7 
           \oplus 0.004 E 
           \oplus  2.1 \left(\frac{E}{100}\right)^{0.3} \, \%,
\end{eqnarray*}
where $E$ is the jet energy in GeV. The four terms in the expression 
respectively represent: the
intrinsic calorimetric resolution; imperfect tracking; leakage and
confusion. This functional form is shown in Figure~\ref{fig:pflow}.
It is worth noting that the predicted jet energy resolutions for
375\,GeV and 500\,GeV jets are in good agreement with those found
for MC events (see Table~\ref{tab:resvsE}); these data were not used
in the determination of the parameterisation of the jet energy resolution.

The ILC jet energy goal of $\sigma_E/E<3.8\,\%$ is reached in the jet energy range
40\,GeV $-$ 420\,GeV. 
Figure~\ref{fig:pflow} also shows a parameterisation of the jet energy resolution 
($\rms$) obtained from a simple sum of the total calorimetric energy deposited in the 
ILD detector concept. It is worth noting that even for the highest energies
jets considered, PFlow reconstruction significantly improves the resolution. 
The performance of PFlow calorimetry is compared to 
${60\,\%}/{\sqrt{E(\mathrm{GeV})}} \oplus 2.0\,\%$ which is intended to give
an {\it indication} of the resolution which might be achieved using a traditional 
calorimetric approach. For a significant range of the jet energies relevant for
the ILC, PFlow results in a jet energy resolution which is roughly a factor
two better than the best at LEP.

\section{Dependence on Hadron Shower Modelling}

The results of the above studies rely on the accuracy of the MC simulation
in describing EM and hadronic showers.
The Geant4 MC provides a good description of
EM showers as has been demonstrated in a series of test-beam 
experiments~\cite{bib:CaliceEM}
using a Silicon-Tungsten ECAL of the type assumed for the ILD detector model.  
However, the uncertainties in the development of hadronic showers are much 
larger~\cite{bib:HSSW}. There are a number of possible effects which 
could affect PFlow performance: 
the hadronic energy resolution;  
the transverse development of hadronic showers which will
affect the performance for higher energy jets where confusion is 
important; and
the longitudinal development of the shower which will affect both the separation 
of hadronic and EM showers and the amount of leakage through 
the rear of the HCAL.

To assess the sensitivity of 
PFlow reconstruction to hadronic shower modelling uncertainties, 
five Geant4 physics lists are compared: 
\begin{itemize}
  \item{\tt QGSP\_BERT}, Quark-Gluon String model\cite{bib:qgs} 
                          with the addition of the Precompound model of nuclear evaporation\cite{bib:precompound} 
                          (QGSP) for high energy interactions, 
                        and the Bertini (BERT) cascade model\cite{bib:bert}
                        for intermediate energy interactions;
  \item{\tt QGS\_BIC},  Quark-Gluon String (QGS)
                        for high energy interactions and the Binary cascade 
                         (BIC) model\cite{bib:bic} for intermediate and low energies;
  \item{\tt FTFP\_BERT}, the Fritiof (FTF) string-based model\cite{bib:fritiof} 
                        with Precompound\cite{bib:precompound} 
                        for high energy interactions and the Bertini cascade 
                        model for intermediate energies;
  \item{\tt LHEP},    based on the Low and High Energy Parameterised modes 
                           (LEP and HEP) of the GHEISHA package\cite{bib:gheisha}
    used in \GEANTthree;
  \item{\tt LCPhys}\cite{bib:lcphys},  which uses a combination of the QGSP, LEP and BERT models.
\end{itemize}
These physics lists represent a wide range of models and result in significantly
different predictions for total energy deposition, and the  
longitudinal and transverse shower profiles. For each Physics list, the calibration
constants in $\PANDORAPFA$ are re-tuned, but no attempt to re-optimise the algorithm
is made. The jet energy resolutions obtained are given in Table~\ref{tab:mcdep}. 
Whilst non-statistical differences are seen,
the rms variations are relatively
small, less than 4.2\,\%. Whilst this might seem surprising, it should be noted 
that the effect on the jet energy resolution of the hadronic modelling
is likely to be predominantly from the 
neutral hadron confusion term. This tends to dilute the sensitivity to
the modelling of hadronic showers. For example, from Table~\ref{tab:recodep} it can 
be seen that if the neutral hadron confusion term for 250\,GeV jets is increased by 25\,\%, 
when added in quadrature to the other terms, the overall jet energy resolution would
only increase by 10\,\%.

\begin{table*}[htb]
\begin{center}
\begin{tabular}{|c|cccc|}
  \hline
  \multicolumn{1}{|l|}{\bf Physics List} & \multicolumn{4}{c|}{\bf\boldmath Jet Energy Resolution $r=\rms(\Ej)/\Ej$}                          \\  
         &    45 GeV           &      100 GeV        &  180 GeV            &    250 GeV          \\ \hline
  {\tt LCPhys}     & $(3.74\pm0.05)\,\%$ & $(2.92\pm0.04)\,\%$ & $(3.00\pm0.04)\,\%$ & $(3.11\pm0.05)\,\%$ \\
  {\tt QGSP\_BERT} & $(3.52\pm0.06)\,\%$ & $(2.95\pm0.06)\,\%$ & $(2.98\pm0.06)\,\%$ & $(3.25\pm0.07)\,\%$ \\
  {\tt QGS\_BIC}  & $(3.51\pm0.06)\,\%$ & $(2.89\pm0.05)\,\%$ & $(3.12\pm0.07)\,\%$ & $(3.20\pm0.07)\,\%$ \\
  {\tt FTFP\_BERT} & $(3.68\pm0.08)\,\%$ & $(3.10\pm0.06)\,\%$ & $(3.24\pm0.06)\,\%$ & $(3.26\pm0.08)\,\%$ \\
  {\tt LHEP}       & $(3.87\pm0.07)\,\%$ & $(3.15\pm0.06)\,\%$ & $(3.16\pm0.06)\,\%$ & $(3.08\pm0.06)\,\%$ \\ \hline
$\chi^2$ (4 d.o.f) &      23.3           &     17.8            &      16.0           &    6.3              \\
  rms/mean ($\sigma_r/\overline{r}$)& 4.2\,\% &     3.9\,\%    &      3.5\,\%        &    2.5\,\%          \\
\hline
\end{tabular}
\caption{Comparison of the jet energy resolution obtained using different hadronic shower
          physics lists. The $\chi^2$ consistency of the different models for each jet energy are
          given as are the rms variations between the five models.
\label{tab:mcdep}}
\end{center}
\end{table*}

From the above study it is concluded that, for $45-250$\,GeV jets, 
the  jet energy resolution obtained from PFlow calorimetry as implemented
in \PANDORAPFA\ does not depend strongly on the hadronic shower
model; the observed differences are less than 5\,\%. This is an important statement;
it argues strongly against the need for a test beam based demonstration of 
PFlow calorimetry (the design of such an experiment would be challenging). 
From test beam data the performance of the ECAL and HCAL systems can be demonstrated
using single particles and the MC can be validated. Once the single
particle performance is demonstrated, the uncertainties in 
extrapolating to the full PFlow performance for jets, which arise from the detailed
modelling of hadronic showers, are likely to be less than $5\,\%$.

\section{Detector Design for Particle Flow Calorimetry}

PFlow calorimetry requires the full reconstruction of
the individual particles from the interaction. The optimisation of
a detector designed for PFlow calorimetry extends 
beyond the calorimeters as tracking information plays a major
role. This section presents a study
of the general features of a detector designed for
high granularity PFlow reconstruction.

\subsection{General Arguments}

PFlow calorimetry requires
the efficient separation of showers from charged hadrons, 
photons and neutral hadrons. This implies high granularity calorimeters
with both the ECAL and HCAL inside the detector solenoid. 
For high energy jets, failures in the ability 
to efficiently separate energy deposits from different particles,
the {\it confusion} term, will dominate the jet energy resolution.
The physical separation of calorimetric energy deposits from different 
particles will be greater in a large detector, scaling as the inner radius
of the ECAL, $R$, in the barrel region and the detector length, $L$, 
in the endcap region. 
There are also arguments favouring a high magnetic field, as this will tend 
to deflect charged particles away from the core of a jet. The scaling law
here is less clear. The separation between a charged particle 
and an {\it initially collinear} neutral particle will scale as $BR^2$. 
However, there is no reason to believe that this will hold on average
for a jet of non-collinear neutral and charged particles. The true dependence 
of PFlow performance on the global detector parameters, $B$ and $R$
has to be evaluated empirically.

\subsection{Particle Flow Optimisation Methodology}

The dependence of the PFlow performance on the main
detector parameters has been investigated using 
\PANDORAPFA.
The studies are based on full 
reconstruction of the tracking and the calorimetric information. 
The results presented here use the
Geant4 simulation of the LDC detector concept~\cite{bib:ldc} which,
from the point of view of PFlow, is essentially the same as the 
ILD detector concept described in Section~\ref{sec:ild}. 
The starting point for the optimisation studies is the LDCPrime detector 
model with a 3.5\,T magnetic field, an ECAL inner radius of 1820\,mm and a 
48 layer (6$\lambda_I$) HCAL. 
The ECAL and HCAL transverse segmentations are $5\times5$\,mm$^2$ 
and $3\times3$\,cm$^2$ respectively. The jet resolution is investigated
as a function of a number of parameters.

\subsection{HCAL Depth}

For good PFlow performance both the ECAL and HCAL need to be within the
detector solenoid. Consequently, in addition to the cost of the HCAL, 
the HCAL thickness impacts the cost of the overall detector through the
radius of the superconducting solenoid. The thickness of the HCAL determines
the average fraction of the jet energy that is contained within the calorimeter system.
The impact of the HCAL thickness on PFlow performance is assessed by 
changing the number of HCAL layers in the LDCPrime model from 32 to 63.
This corresponds to a variation of $4.0-7.9$\,$\lambda_I$
in the HCAL ($4.8-8.7$\,$\lambda_I$ in the ECAL+HCAL combined).

\begin{figure}[htb]
\epsfxsize=1.0\columnwidth
\centerline{\includegraphics[width=1.0\columnwidth]{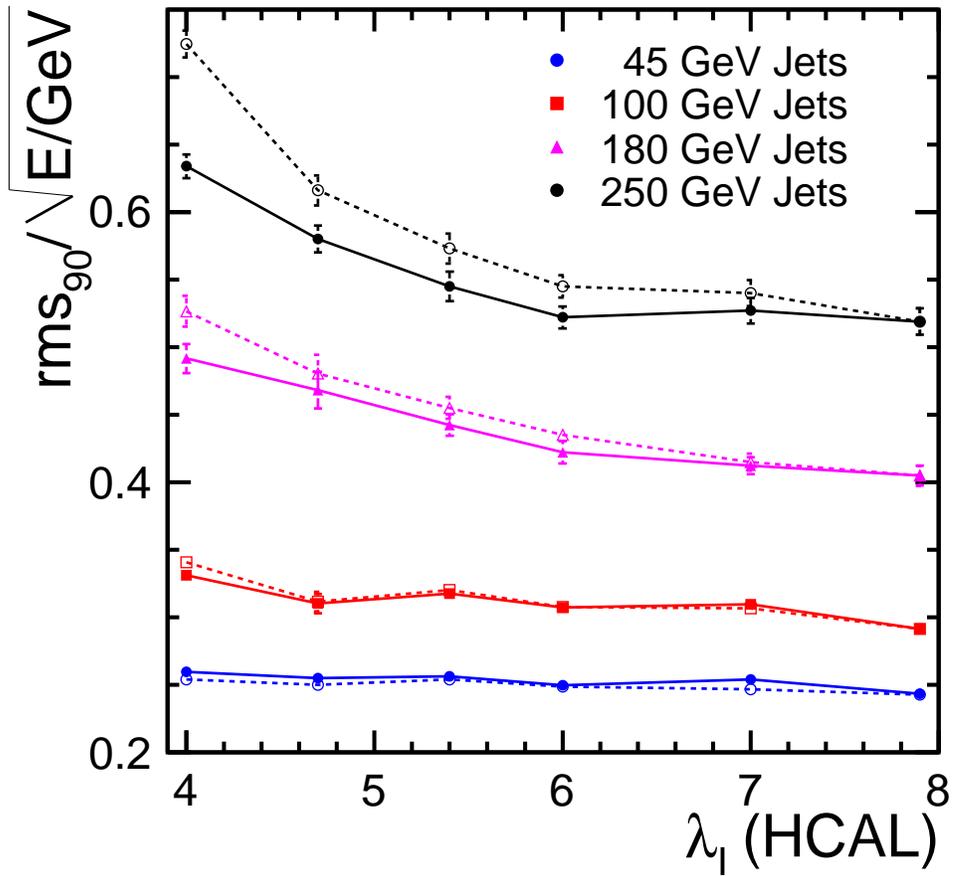}}
 \caption{Jet energy resolutions ($\rmsn$) for the LDCPrime as a function of
          the thickness (normal incidence) of the HCAL. In addition, the ECAL contributes
          $0.8\,\lambda_I$. Results are shown with (solid markers) and without (open 
          markers) taking into account
          energy depositions in the muon chambers. All results are based on 
         $\Zzero\rightarrow\uubar,\ddbar,\ssbar$ with generated polar angle in the
         barrel region of the detector,  $|\cos\theta_{\qq}|<0.7$.
  \label{fig:pfa_hcal}}
\end{figure}

The study of the optimal HCAL thickness depends on the possible
use of the instrumented return yoke (the muon system) to correct for 
leakage of high energy showers out of the rear of the HCAL. The 
effectiveness of this approach is limited by the fact that,
for much of the polar angle, the muon system is behind the relatively
thick solenoid ($2\lambda_I$ in the \MOKKA\ simulation of the detector). 
Nevertheless, to assess the possible impact of using the
muon detector as a ``tail-catcher'', the energy depositions in the muon 
detectors were included in the \PANDORAPFA\ reconstruction. Whilst 
the treatment could be improved upon, it provides an indication 
of how much of the degradation in jet energy resolution 
due to leakage can be recovered in this way. The results are 
summarised in Figure~\ref{fig:pfa_hcal} which shows the jet energy resolution 
obtained from \PANDORAPFA\ as a function of the HCAL thickness.
The effect of leakage is
clearly visible, with about half of the degradation in resolution being recovered
when including the muon detector information. For jet energies of 100\,GeV or less,
leakage is not a major contributor to the jet energy resolution provided the 
HCAL is approximately $4.7\lambda_I$ thick (38 layers). However, for 
$180-250$\,GeV jets this is not sufficient; for leakage not to contribute
significantly to the jet energy resolution at $\roots=1$\,TeV, the results in 
Figure~\ref{fig:pfa_hcal} suggest that the HCAL thickness should be 
between $5.5-6.0\lambda_I$ for an ILC detector.

\subsection{Magnetic Field versus Detector Radius}

The LDCPrime model assumes a magnetic field of
3.5\,T and an ECAL inner radius of 1820\,mm. A number of variations 
on these parameters were studied: 
i) variations in the ECAL inner 
radius from $1280-2020$\,mm
with $B=3.5$\,T; ii) variations the $B$ from $2.5-4.5$\,T with 
$\rECAL=1825$\,mm; and iii) variations of both $B$ and $\rECAL$. 
In total thirteen sets of parameters 
were considered spanning a wide range of
$B$ and $\rECAL$. 
The parameters include those considered by the LDC, GLD~\cite{bib:gld}, and SiD~\cite{bib:sid} 
detector concept groups for the ILC. In each case PFlow performance was
evaluated for 45, 100, 180, and 500\,GeV jets. 

\begin{figure*}
\includegraphics[width=0.5\columnwidth]{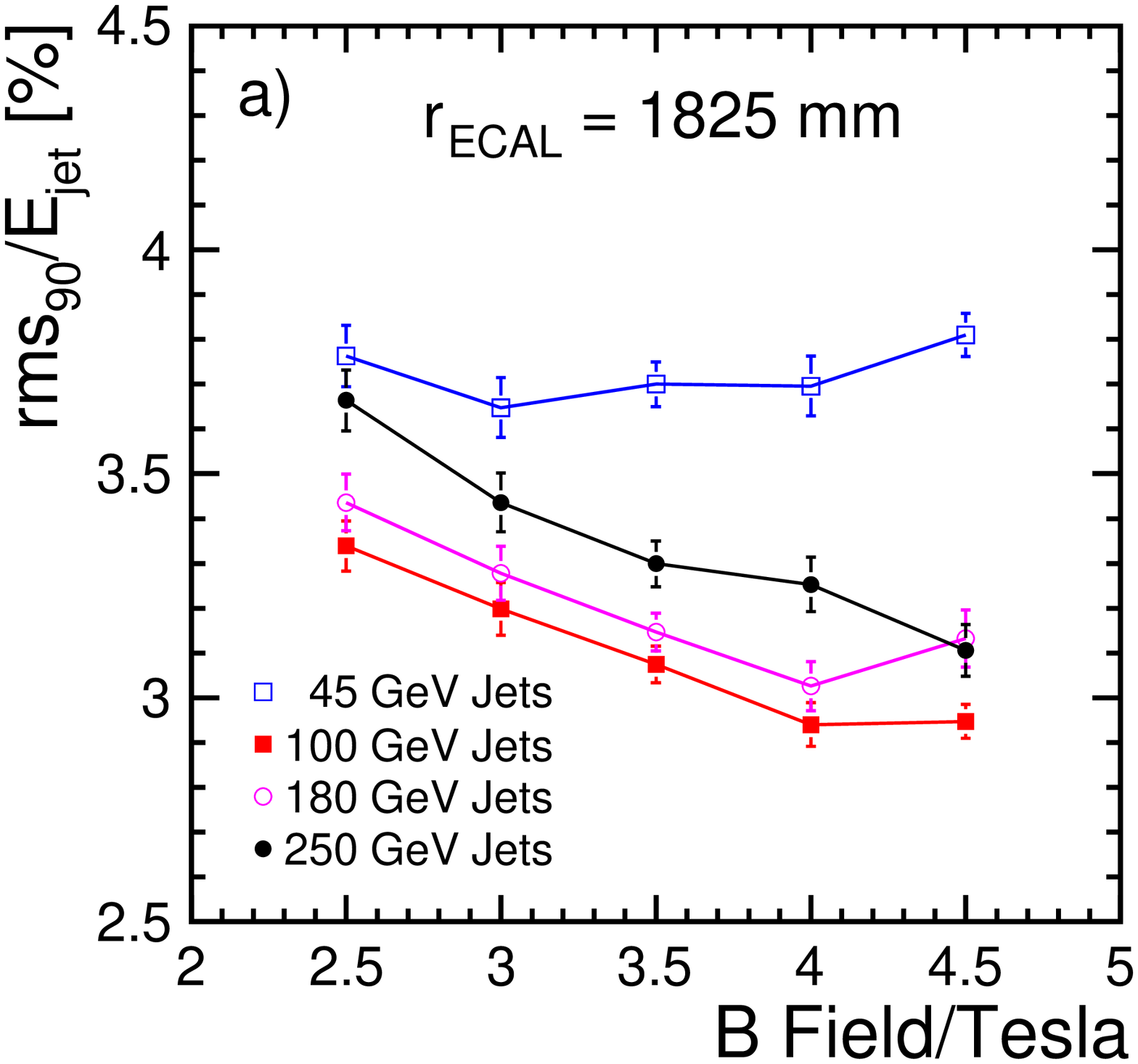}
\includegraphics[width=0.5\columnwidth]{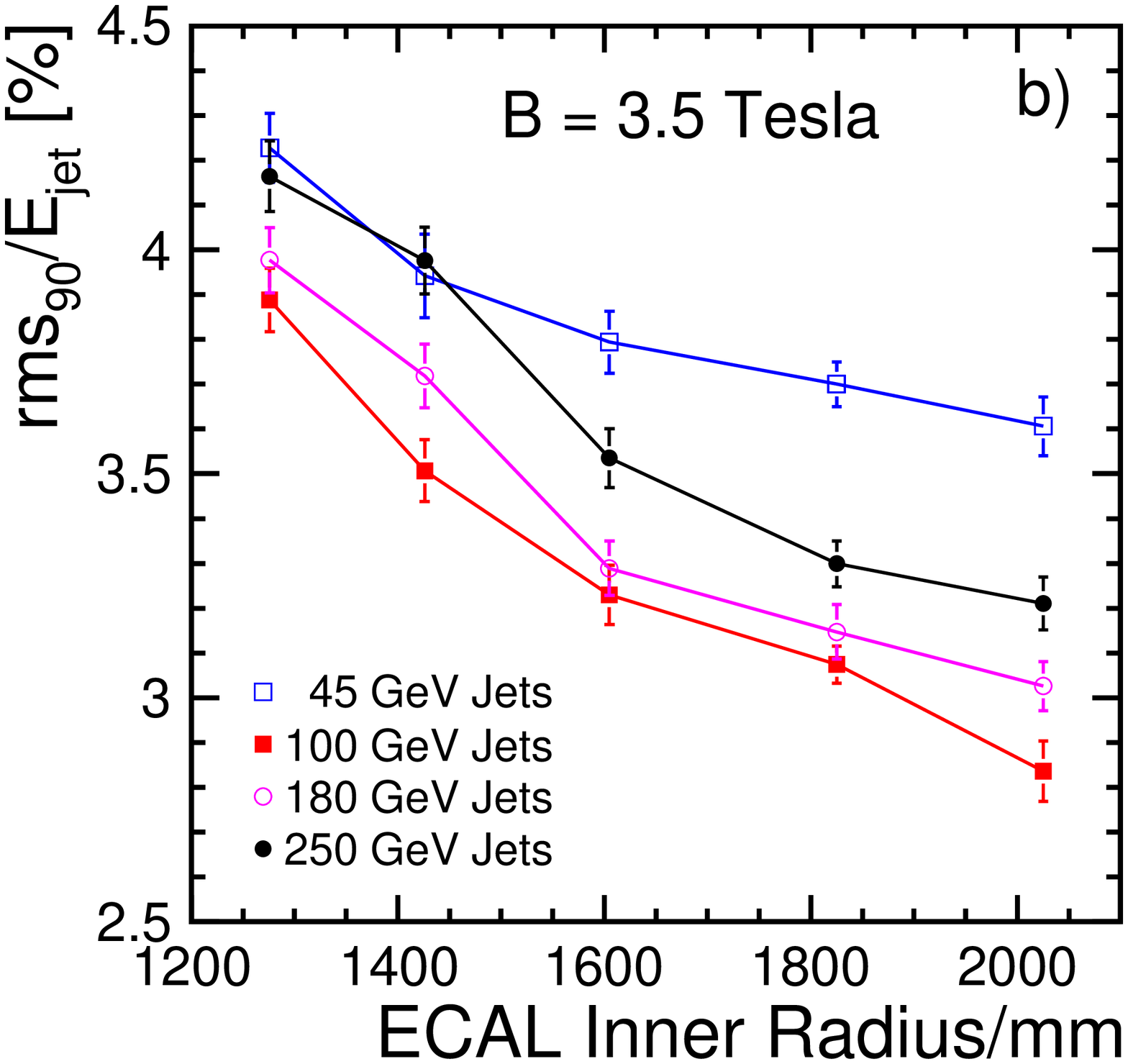}
 \caption{{\bf a)} the dependence of the jet energy resolution ($\rmsn$) on 
             the magnetic field for a fixed ECAL inner radius. {\bf b)} 
               the dependence of the jet energy resolution ($\rmsn$) on the ECAL
               inner radius a fixed value of the magnetic field.
             The resolutions are obtained from $\Zzero\rightarrow \uubar, \ddbar, \ssbar$ decays 
             at rest. The errors shown are statistical only.
 \label{fig:pfa_b_versus_r}}
\end{figure*}

Figure~\ref{fig:pfa_b_versus_r} shows the dependence of
the jet energy resolution 
as a function of: a) magnetic field (fixed $\rECAL$)
and b) ECAL inner radius (fixed $B$). For 45\,GeV jets, the
dependence of the jet energy resolution on $B$ and $\rECAL$ is rather
weak because, for these energies, it is the intrinsic calorimetric
energy resolution rather than the confusion term that dominates. For
higher energy jets, where the confusion term dominates the resolution,
the jet energy resolution shows a stronger dependence on $\rECAL$ 
than $B$.

The jet energy resolutions are reasonably well described by the function:
\begin{eqnarray*}
   &&  \frac{\rms}{E}  =  \frac{21}{\sqrt{E}} 
           \oplus 0.7 
           \oplus 0.004 E \\
   && \ \ \ \  \ \ \ \ \ \ \ \oplus \ 2.1 
                   \left(\frac{\rECAL}{1825}\right)^{-1.0}
                   \left(\frac{B}{3.5}\right)^{-0.3}
                   \left(\frac{E}{100}\right)^{0.3} \, \%,
\end{eqnarray*}
where $E$ is measured in GeV, $B$ in Tesla, and $\rECAL$ in mm.
This is the quadrature sum of four terms: 
i) the estimated contribution to the jet energy resolution from the intrinsic calorimetric resolution; 
ii) the contribution from track reconstruction;
iii) the contribution from leakage; and 
iv) the contribution from the confusion term obtained empirically from a fit to the data of 
      Figure~\ref{fig:pfa_b_versus_r} and several models where both $B$ and $\rECAL$ are
      varied~\cite{bib:ildloi}. 
In fitting the confusion term, a power-law form, $\kappa B^\alpha \rECAL^\beta E^\gamma$, is assumed. 
This functional form provides a reasonable parameterisation of the data; 
the majority of the data points lie within 2$\sigma$ of the parameterisation.

From the perspective of the optimisation of a  detector for PFlow, these studies show that 
for the \PANDORAPFA\ algorithm, the confusion
term scales as approximately $B^{0.3}R$, {\it i.e.} for good PFlow
performance a large detector radius is significantly more important 
than a very high  magnetic field. 

\subsection{ECAL and HCAL Design}

\label{sec:calseg}

\begin{figure*}
\includegraphics[width=0.5\columnwidth]{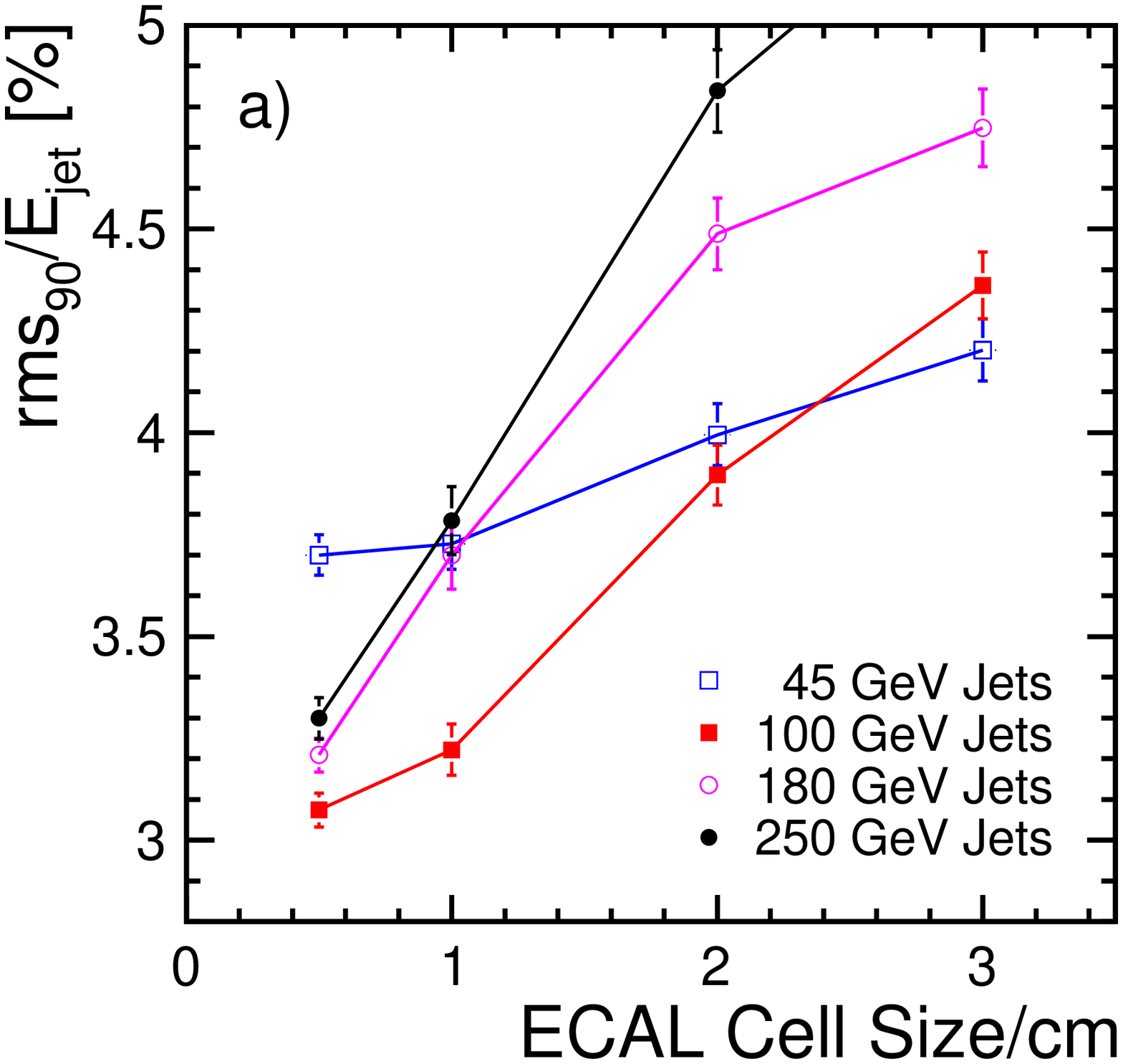}
\includegraphics[width=0.5\columnwidth]{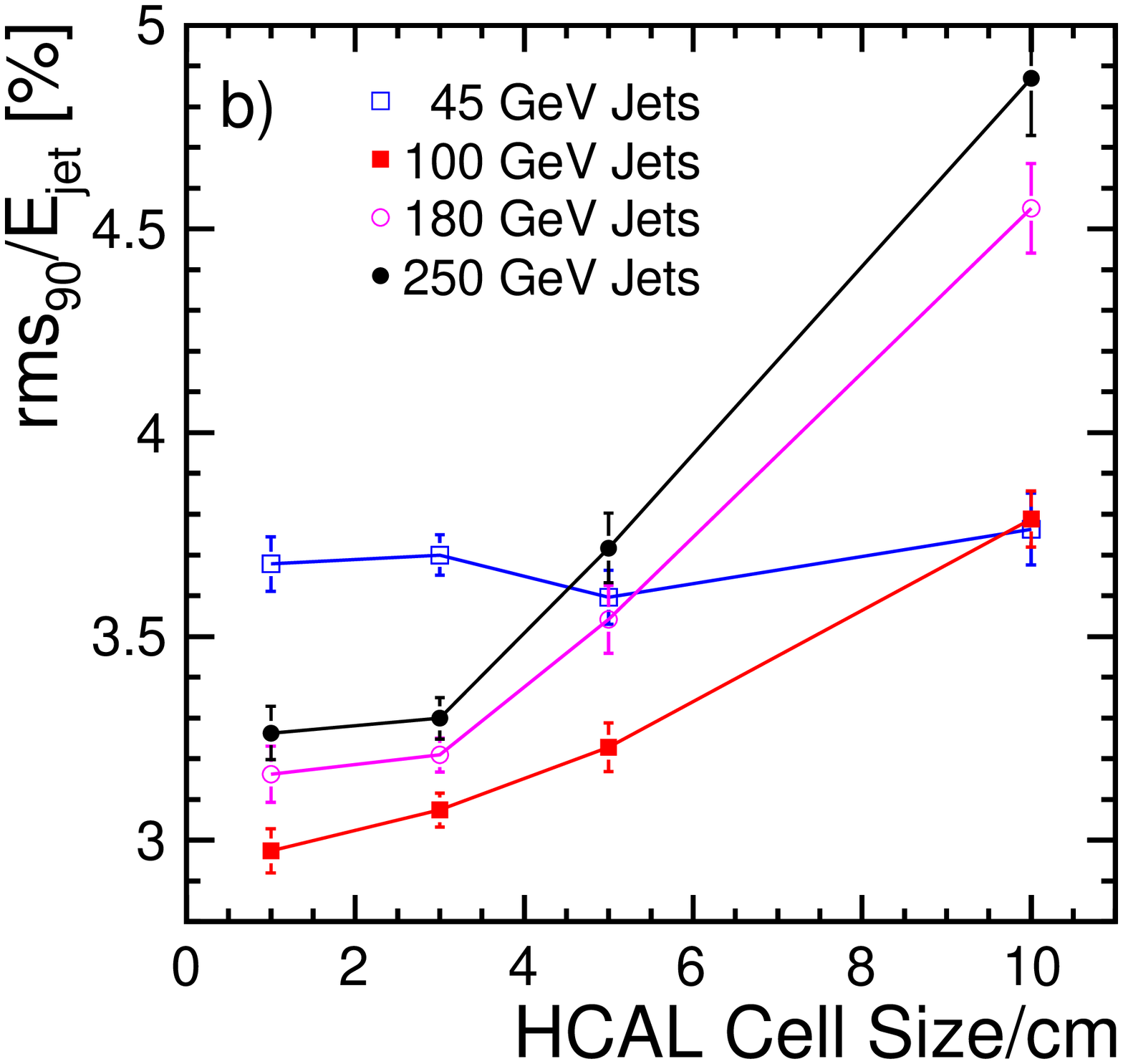}
 \caption{{\bf a)} the dependence of the jet energy resolution ($\rmsn$) on 
             the ECAL transverse segmentation (Silicon pixel size)  
             in the LDCPrime model. {\bf b)} 
               the dependence of the jet energy resolution ($\rmsn$) on the HCAL
              transverse segmentation (scintillator tile size)  
             in the LDCPrime model.  
             The resolutions are obtained from $\Zzero\rightarrow \uubar, \ddbar, \ssbar$ decays 
             at rest. The errors shown are statistical only.
 \label{fig:pfa_segmentation}}
\end{figure*}

The dependence of PFlow performance on the transverse segmentation of the
ECAL was studied using modified versions of the LDCPrime model. The jet energy resolution 
is determined for different ECAL
Silicon pixel sizes; $5\times5$\,mm$^2$, $10\times10$\,mm$^2$,
$20\times20$\,mm$^2$, and $30\times30$\,mm$^2$. The two main clustering
parameters in the \PANDORAPFA\ algorithm were re-optimised for each ECAL 
granularity. The PFlow performance results are summarised in 
Figure~\ref{fig:pfa_segmentation}a. For 45\,GeV jets, the dependence is relatively
weak since the confusion term is not the dominant contribution to the resolution.
For higher energy jets, a significant degradation in performance is observed
with increasing pixel size. Within the context of the current
reconstruction, the ECAL
transverse segmentations have to be at least as fine as $10\times10$\,mm$^2$ to
meet the ILC jet energy requirement of $\sigma_E/E<3.8\,\%$ for the
jet energies relevant at $\roots = 1$\,TeV, with $5\times5$\,mm$^2$ 
being preferred. 

A similar study was performed for the HCAL. The jet energy resolution
obtained from \PANDORAPFA\ was investigated for HCAL 
scintillator tile sizes of $1\times1$\,cm$^2$, $3\times3$\,cm$^2$,
$5\times5$\,cm$^2$ and $10\times10$\,cm$^2$. 
The PFlow performance results are summarised in 
Figure~\ref{fig:pfa_segmentation}b. From this study, it is concluded
that the ILC jet energy resolution goals can be achieved an HCAL
transverse segmentation of $5\times5$\,cm$^2$. For higher energy
jets going to $3\times3$\,cm$^2$ leads to a significant improvement
in resolution. From this study there appears to be no significant
motivation for $1\times1$\,cm$^2$ granularity over $3\times3$\,cm$^2$.
The results quoted here are for an analogue scintillator tile calorimeter.
The conclusions for a digital, {\it e.g.} RPC-based, HCAL
might be different.

\subsection{Summary}

Based on the above studies, the general features of a detector 
designed for high granularity PFlow calorimetry are:
\begin{itemize}
 \item ECAL and HCAL should be inside the solenoid.
 \item The detector radius should be as large as possible, the confusion term scales approximately with
         the ECAL inner radius as  $R^{-1}$.
 \item To fully exploit the potential of PFlow calorimetry the 
         ECAL transverse segmentation should be at least as fine as $5\times5$\,mm$^2$.
 \item For the HCAL longitudinal segmentation considered here, there
        is little advantage in transverse segmentation finer than $3\times3$\,cm$^2$.
 \item The argument for a very high magnetic field is relatively weak as the confusion term
        scales as $B^{-0.3}$.
\end{itemize}
These studies, based on the \PANDORAPFA\ algorithm, motivated the design 
of the ILD detector concept for the ILC as is discussed in more detail 
in Chapter~2 of~\citep{bib:ildloi}.

\section{Particle Flow for Multi-TeV Colliders}

In this section the potential of PFlow
Calorimetry at a multi-TeV $\epem$ collider, such as CLIC~\cite{bib:clic}, 
is considered. Before the results from the LHC are known it is difficult to
fully define the jet energy requirements for a CLIC detector. However,
if CLIC is built, it is likely that the construction will be phased with initial 
operation at ILC-like energies followed by high energy operation 
at $\roots\sim3$\,TeV. It has been shown in this paper that PFlow calorimetry
is extremely powerful for ILC energies. Given that the confusion term increases
with energy, it is not {\it a priori} clear that PFlow calorimetry is suitable
for higher energies. This questions needs to be considered in the context
of the possible physics measurements where jet energy resolution is 
likely to be important at $\roots\sim3$\,TeV. For example, the reconstruction
of the jet energies in $\epem\rightarrow\qq$ events is unlikely to be interest. 
Assuming the main physics processes of interest consist of final states with 
between six and eight fermions, the likely 
relevant jet energies will be in the range $375-500$\,\GeV. 
To study the potential of the PFlow calorimetry for these jet energies the ILD concept, 
which is optimised for ILC energies, was modified; the HCAL thickness was increased
from $6\,\lambda_I$ to $8\,\lambda_I$ and the magnetic field was increased from 
3.5\,T to 4.0\,T. The jet energy resolution obtained for jets from 
 $\Zzero\rightarrow \uubar, \ddbar, \ssbar$
decays at rest are listed in Table~\ref{tab:resvsEclic}. For high energy jets,
the effect of the increased HCAL thickness (the dominant effect) and increased magnetic
field is significant. Despite the increased particle densities, the jet energy 
resolution $(\rms)$ for 500\,GeV jets obtained from PFlow is 3.5\,\%. This is
equivalent to $78\,\%/\sqrt{E(\mathrm{GeV})}$. This is likely to be 
{\it at least} competitive with a traditional calorimetric approach,
particularly when the constant term in Equation~\ref{eqn:cal} 
and the contribution from non-containment are accounted for. 
Furthermore, it should be remembered that \PANDORAPFA\ has
not been optimised for such high energy jets and improvements can be expected.
It is also worth noting that the purely calorimetric energy resolution ($\rms$)
for 500\,GeV jets with the modified ILD concept is equivalent to 
$115\,\%/\sqrt{E(\mathrm{GeV})}$ and, thus, the gain from PFlow reconstruction is
still significant.

\begin{table*}[thb]
\begin{center}
\begin{tabular}{|r|cc|cc|}
\hline
  {\bf Jet Energy}        &  \multicolumn{2}{|c}{\bf\boldmath $\rms(\Ejj)/\sqrt{\Ejj}$} &  \multicolumn{2}{|c|}{\bf\boldmath $\rms(\Ej)/\Ej$ }   \\ 
                          &  3.5\,T \& 6\,$\lambda_I$  & 4\,T \& 8\,$\lambda_I$      & 3.5\,T \& 6\,$\lambda_I$  & 4\,T \& $8\,\lambda_I$   \\ \hline
  45 GeV                  &  25.2\,\% & 25.2\,\%   &   $(3.74\pm0.05)\,\%$   & $(3.74\pm0.05)\,\%$  \\
  100 GeV                 &  29.2\,\% & 28.7\,\%   &   $(2.92\pm0.04)\,\%$   & $(2.87\pm0.04)\,\%$  \\
  180 GeV                 &  40.3\,\% & 37.5\,\%   &   $(3.00\pm0.04)\,\%$   & $(2.80\pm0.04)\,\%$  \\
  250 GeV                 &  49.3\,\% & 44.7\,\%   &   $(3.11\pm0.05)\,\%$   & $(2.83\pm0.05)\,\%$  \\
  375 GeV                 &  81.4\,\% & 71.7\,\%   &   $(3.64\pm0.05)\,\%$   & $(3.21\pm0.05)\,\%$  \\
  500 GeV                 &  91.6\,\% & 78.0\,\%   &   $(4.09\pm0.07)\,\%$   & $(3.49\pm0.07)\,\%$  \\ \hline
\end{tabular}
\caption{Comparisons of jet energy resolutions for two sets of detector parameters. 
            This jet energy resolution shown is for $(\Zzero/\gamma)^*\rightarrow$uds events 
            with $|\cosqq|<0.7$.  
            It is expressed as: i) the effective
            constant $\alpha$ in $\rms(\Ejj)/\Ejj = \alpha(\Ejj)/\sqrt{\EjjGeV}$, where
            $\Ejj$ is the total reconstructed energy; and ii) the
            fractional jet energy resolution for a single jet where $\rms(\Ej)=\rms(\Ejj)/\sqrt{2}$.
\label{tab:resvsEclic}}
\end{center}
\end{table*}

\subsection{Gauge Boson Mass Reconstruction}

\label{sec:massreco}

\begin{figure}[hbt]
\epsfxsize=1.0\columnwidth
\centerline{\includegraphics[width=1.0\columnwidth]{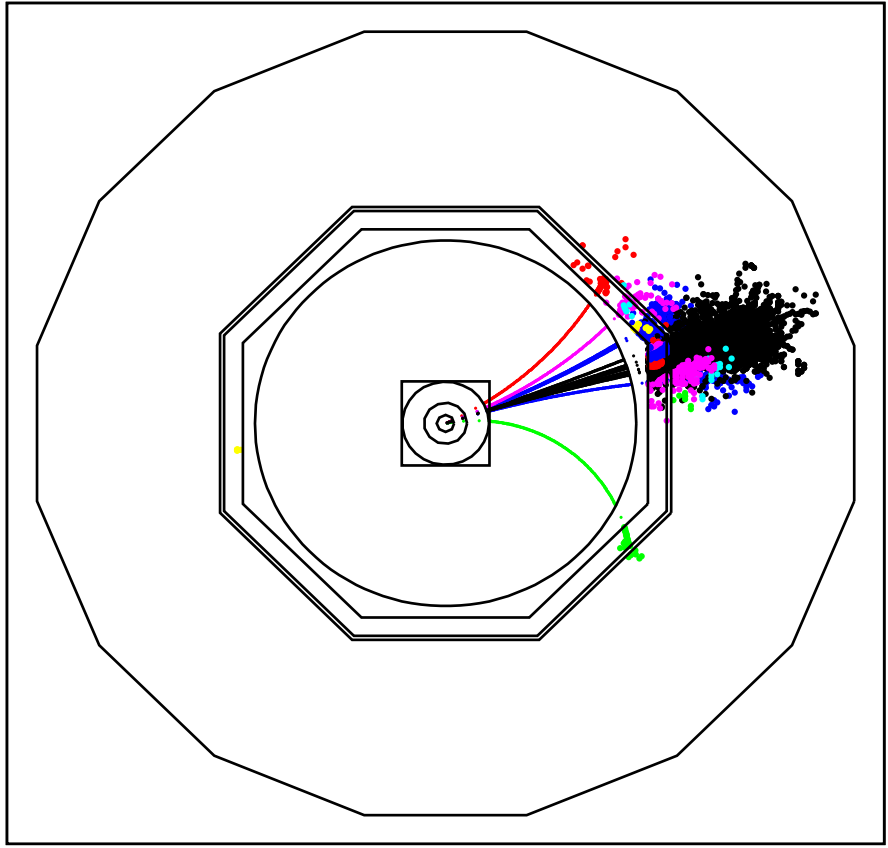}}
 \caption{An example of a $\Zzero\rightarrow\ddbar$ decay with $E_{Z}=1$\,TeV
          produced in a simulated $\epem\rightarrow\Zzero\Zzero\rightarrow\nu\overline{\nu}\ddbar$
          interaction in the ILD detector concept.
  \label{fig:monojet}}
\end{figure}

A requirement for a detector at a future linear collider 
is the ability to separate hadronic $\Wboson$ and $\Zzero$ decays.
It was on this basis that the ILC jet energy resolution goal of
$\sigma_E/E \lesssim 3.8\,\%$ was justified. 
The performance of PFlow calorimetry has, up to this point, been
considered in terms of the jet energy resolution from particles 
decaying at rest. This is reasonable since one of the main goals 
of a future linear collider will be to study the
physics Beyond the Standard Model (BSM) which hopefully will be
uncovered at the LHC. Thus, many of the processes of interest are 
likely to be produced relatively close to threshold. 
In this case, the new particle(s) will be produced almost at rest.
Similarly, for processes where a new particle is produced in 
association with one or more gauge bosons, the gauge bosons will be produced 
almost at rest. However, it is also possible that gauge bosons may be 
produced from the decays of BSM particles. In this case, 
the $\Wboson$/$\Zzero$  decays will not be at rest and the di-jet system will 
be boosted. At a multi-TeV lepton collider the boost may be significant as the 
energies of the gauge bosons are potentially in the range 
500\,GeV$-$1\,TeV. For PFlow calorimetry there are a number of 
effects associated with highly boosted jets:
\begin{itemize}
 \item The jet particle multiplicities are lower than those for jets of the same 
    energy produced from decays at rest. This increases the average energy 
    of the particles in the jet and, consequently, will result in 
    less containment of the hadronic showers (greater leakage);
 \item The energies of the jets in the di-jet system will, in general, not be equal. 
    Where
    one of the jets is much higher in energy than the other PFlow performance will
    tend to degrade.  
 \item The high jet boost decreases the average 
     separation of the particles in the jet. This will tend to 
      increase the
     confusion term. 
 \item The two jets from the decay of a highly boosted gauge 
     boson will tend to overlap to form a ``mono-jet'', as shown
     in Figure~\ref{fig:monojet}.
     The overlapping of jets has the potential to increase the confusion term. 
\end{itemize}
Due to the likely increased confusion term, reconstructing the invariant 
mass of high energy gauge bosons presents a challenge for PFlow calorimetry. 
However, it should be noted that it may be even more challenging for a traditional 
calorimetric approach as it is now necessary to reconstruct the invariant mass of a
single system of nearby particles which will not be well-resolved in the calorimeters.

\begin{figure}[hbtp]
\epsfxsize=1.0\columnwidth
\centerline{\includegraphics[width=1.0\columnwidth]{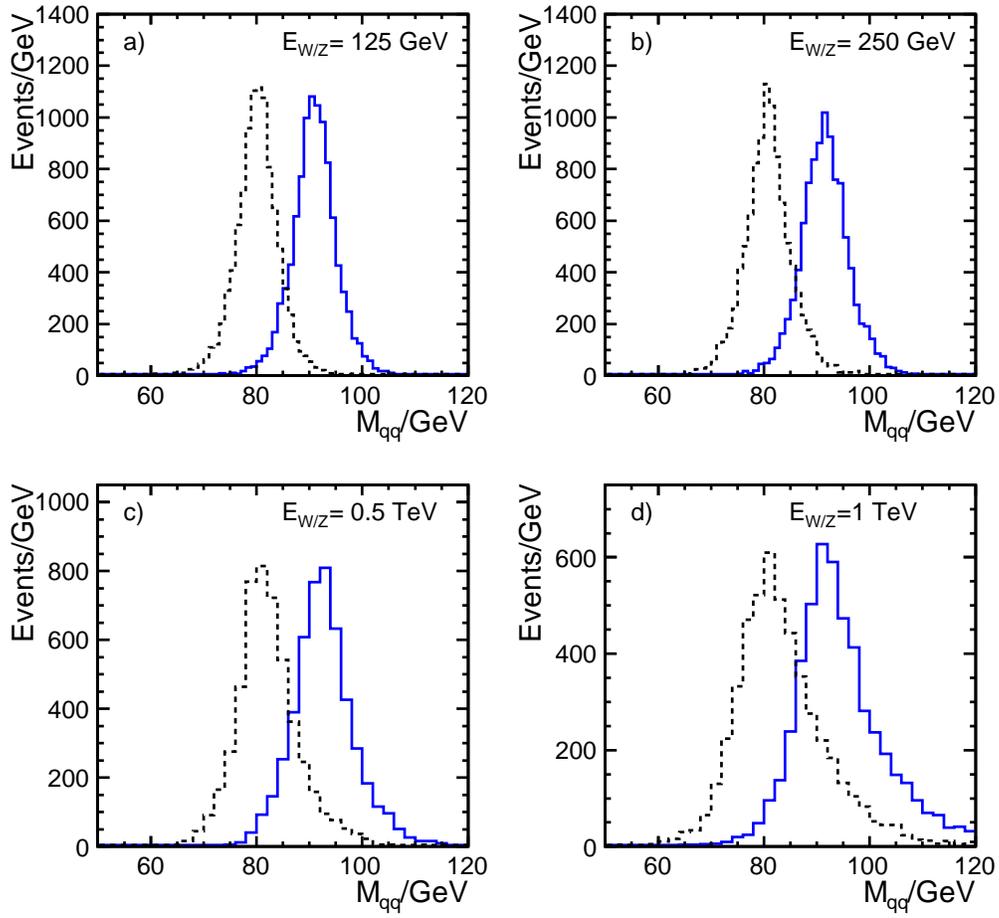}}
 \caption{Reconstructed invariant mass distributions for the hadronic system in simulated
         $\Zzero\Zzero\rightarrow\ddbar\nu\overline{\nu}$ and
         $\WW\rightarrow \mathrm{u}\overline{\mathrm{d}}\mu^-\overline{\nu}_\mu$ events 
         as simulated in the modified ILD detector model.  
  \label{fig:massreco}}
\end{figure}

The PFlow reconstruction of boosted gauge bosons has been investigated by generating
MC samples of $\Zzero\Zzero\rightarrow\ddbar\nu\overline{\nu}$ and
$\WW\rightarrow \mathrm{u}\overline{\mathrm{d}}\mu^-\overline{\nu}_\mu$ events at
$\roots=$ 0.25, 0.5, 1.0 and 2.0\,TeV. These final states give
clean samples of single hadronic $\Zzero$ and $\Wboson$ decays (the
muons from the $\Wboson$ decays are easy to identify and remove). The PFlow reconstructed
$\Wboson$ and $\Zzero$ invariant mass distributions are shown in 
Figure~\ref{fig:massreco} and the corresponding mass resolutions are given
in Table~\ref{tab:massreco}. A direct comparison with
the jet energy resolutions of Table~\ref{tab:resvsEclic} is not straightforward
due to the effects described above. However, the mass resolution 
$(\rms)$ of 2.8\,GeV obtained from decays of gauge bosons with $E=125$\,GeV 
is compatible with that 
expected from the jet energy resolution of Table~\ref{tab:resvsEclic}
after accounting for the gauge boson width. 

 \begin{table}[thb]
\begin{center}
\begin{tabular}{|r|cccc|}
\hline
  \multicolumn{1}{|c|}{\bf\boldmath $E_{W/Z}$} & {\bf\boldmath $\rms(m)$}  & {\bf\boldmath $\sigma_m/m$} & {\bf\boldmath W/Z sep}  & {\bf\boldmath $\epsilon$} \\ \hline
  125\,GeV        &  2.8\,GeV   & 2.9\,\%  & $2.7\sigma$   & 91\,\%  \\ 
  250\,GeV        &  3.0\,GeV   & 3.5\,\%  & $2.5\sigma$   & 89\,\%  \\ 
  500\,GeV        &  3.9\,GeV   & 5.1\,\%  & $2.1\sigma$   & 84\,\%  \\ 
 1000\,GeV        &  6.4\,GeV   & 7.0\,\%  & $1.5\sigma$   & 78\,\%  \\ 
\hline 
\end{tabular}
\caption{Invariant mass resolutions for the hadronic system in simulated
         $\Zzero\Zzero\rightarrow\ddbar\nu\overline{\nu}$ and
         $\WW\rightarrow \mathrm{u}\overline{\mathrm{d}}\mu^-\overline{\nu}_\mu$ events in the
         ILD detector concept. The $\Wboson$/$\Zzero$ separation numbers, which take into account the
         tails, are defined such that a $2\sigma$ separation means that the optimal cut 
         in the invariant mass distribution results in 15.8\,\% of events being mis-identified. 
         The equivalent $\Wboson/\Zzero$ identification efficiencies, $\epsilon$, are 
         given in the final column. Even with infinitely good
         mass resolution, the best that can be achieved is 94\,\% due to the tails of
         the Breit-Wigner distribution and, thus, the possible range for 
         $\epsilon$ is $50-94\,\%$; 
\label{tab:massreco}}
\end{center}
\end{table}

For the ILC operating at $\roots = 0.5-1.0$\,TeV, the typical energies of the
gauge bosons of interest are likely to be in the range 
$E_{W/Z} = 125-250$\,GeV.  Here the 
reconstructed $\Wboson$ and $\Zzero$ mass peaks are well resolved.
The statistical separation, 
which is quantified in Table~\ref{tab:massreco}, is approximately $2.5\sigma$, {\em i.e.}
the separation between the two peaks is approximately 2.5 times greater the 
effective mass resolution. 

For CLIC operating at $\roots=3$\,TeV, the relevant gauge 
boson energies are likely
to be in the range $0.5-1.0$\,TeV. At the low end of this range
there is reasonable separation ($2.1\sigma$) between the $\Wboson$ and $\Zzero$ peaks.
Even for 1\,TeV $\Wboson$/$\Zzero$ decays, where the events mostly appear as
a single energetic mono-jet, the mass resolution achieved by the current 
version of \PANDORAPFA\ allows separation between $\Wboson$ and $\Zzero$ 
decays at the $1.5\sigma$ level. It should be remembered that \PANDORAPFA\ has
not been optimised for such high energy jets, and these results represent
a lower bound on what can be achieved. 
From this result it is concluded that PFlow calorimetry is certainly not ruled
out for a multi-TeV lepton collider.

\section{Conclusions}  

A sophisticated particle flow reconstruction algorithm, \PANDORAPFA, has been
developed to study the potential of high granularity Particle Flow calorimetry 
at a future linear collider. The algorithm incorporates a number of techniques,
{\it e.g.} topological clustering and statistical reclustering, which take advantage
of the highly segmented calorimeters being considered for the ILC and beyond.

\PANDORAPFA\ has been applied to the reconstruction
of simulated events in the ILD detector concept for the ILC. The results presented in this
paper provide the first conclusive demonstration that Particle Flow Calorimetry
can meet the ILC requirements for jet energy resolution. For jets in the 
energy range $40-400$\,GeV, the jet energy resolution, $\sigma_E/E$, is better
than 3.8\,\%. For the jet energies relevant at the ILC, the jet energy resolution
is approximately a factor of two better than the best achieved at LEP. The conclusions
do not depend strongly on the details of the modelling of hadronic showers. 

\PANDORAPFA\ has been used to investigate the factors limiting the performance of Particle Flow
calorimetry. For jet energies below approximately 100\,\GeV, the intrinsic calorimetric 
resolution dominates the jet energy resolution. For higher energy jets, the confusion term
({\it i.e.} imperfect reconstruction) dominates. The largest single contribution to the
confusion term arises from the mis-assignment of energy from neutral hadrons.

\PANDORAPFA\ has been used to study design of a detector optimised for
high granularity Particle Flow calorimetry demonstrating the importance of
high transverse segmentation in the electromagnetic and hadron calorimeters.
The confusion term, which dominates the jet energy resolution for high energy
jets, scales as approximately $B^{-0.3}\rECAL^{-1}$, where $B$ is the solenoidal 
magnetic field strength and $\rECAL$ is the inner radius of the electromagnetic
calorimeter. 

In addition, \PANDORAPFA\ has been used to perform a preliminary study of
the potential of Particle Flow calorimetry at a multi-TeV collider such
as CLIC. For 
decays at rest, a jet energy resolution below $3.8\,\%$ is achievable for
jets with energies less than approximately 600\,GeV. Reasonable separation 
of the hadronic decays of $\Wboson$ and $\Zzero$ bosons is achievable for
$\Wboson$/$\Zzero$ energies of up to approximately 1\,TeV. 

In conclusion, the studies described in this paper provide the first proof 
of principle of Particle Flow calorimetry at a future lepton collider. For 
ILC energies, $\roots=0.5-1.0$\,TeV, unprecedented jet energy resolution 
can be achieved. Whilst the potential at a multi-TeV collider needs further 
investigation, the results presented in this paper are promising.

\section{Acknowledgements}  

I would like to acknowledge: the UK Science and Technology Facilities Council (STFC)
for the continued support of this work; my colleagues on the ILD detector concept
for providing the high quality simulation and software frameworks used for these
studies; Vasily Morgunov for many interesting and useful discussions on Particle 
Flow reconstruction; and David Ward for reading the near final draft of this paper.


\begin{footnotesize}

\end{footnotesize}

\newpage




\end{document}